\documentclass[11pt]{article}
\usepackage[utf8x]{inputenc}
\usepackage[T1]{fontenc}
\usepackage[english]{babel}
\usepackage{amsmath,amssymb}
\usepackage{graphicx}
\usepackage{xspace}
\usepackage{cite}

\usepackage[a4paper,nohead]{geometry}

\usepackage[bookmarks=true,bookmarksopen=true,colorlinks=true,breaklinks=true,linkcolor=blue,citecolor=blue]{hyperref}

\usepackage[font=small,format=plain,labelfont=bf,up,textfont=it,up]{caption} 
\usepackage[affil-it,max2]{authblk}

\usepackage[titletoc]{appendix}
\hypersetup{%                                                                
  pdftitle = {Criticality of inhomogeneous Nariai-like cosmological models},
  pdfsubject = {},
  pdfauthor = {F. Beyer, L. Escobar, J. Frauendiener},
  pdfkeywords = {}%                                                             
}

\newtheorem{result}{Result}
\newcommand{\Eqref}[1]{Eq.~\eqref{#1}}
\newcommand{\Eqsref}[1]{Eqs.~\eqref{#1}}
\newcommand{\Sectionref}[1]{Section~\ref{#1}}
\newcommand{\Appendixref}[1]{Appendix~\ref{#1}}

\newcommand{\Figref}[1]{Fig.~\ref{#1}}
\newcommand{\Figsref}[1]{Figs.~\ref{#1}}
\newcommand{\s}{\hspace{0.1cm}}

\newcommand{\R}{\ensuremath{\mathbb R}\xspace}

\newcommand{\So}{\ensuremath{\mathbb{S}^1}\xspace}
\newcommand{\St}{\ensuremath{\mathbb{S}^2}\xspace}
\newcommand{\Sth}{\ensuremath{\mathbb{S}^3}\xspace}
\newcommand{\U}{\ensuremath{\mathrm{U(1)}}\xspace}

\newcommand{\mbar}{\overline{m}}

\newcommand{\keyword}[1]{\emph{#1}}
\newcommand{\df} {\mathrm{d}}

\title{Criticality of inhomogeneous Nariai-like cosmological models}
\author[]{F. Beyer\footnote{Email: fbeyer@maths.otago.ac.nz.} }
\author[]{L. Escobar\footnote{Email: lescobar@maths.otago.ac.nz.} }
\author[]{J. Frauendiener \footnote{Email: joergf@maths.otago.ac.nz.} }
\affil[]{Department of Mathematics and Statistics, University of Otago,  New Zealand.}
\date{}

\numberwithin{equation}{section}
\graphicspath{{.}{plots/}}

\begin{document}

\maketitle

%%%%%%%%%%%%%%%%%%%%%%%%%%%%%%%%%%%%%%%%%%%%%%%%%%%%%%%%%%%%%%%%%%%%%%%%%%%%%%%%%%%%%%%%%%%%%%%%%%%%%%%%%%%%%%%%%%%%%%%%%%%%%%%%%
\begin{abstract}
In this paper, we construct and study solutions of Einstein's
equations in vacuum with a positive cosmological constant which can be
considered as inhomogeneous generalizations of the Nariai
cosmological model. Similar to this Nariai spacetime, our solutions are at the
borderline between gravitational collapse and de-Sitter-like
exponential expansion. Our studies focus in particular
 on the intriguing oscillatory dynamics which we discover. Our investigations are
 carried out both analytically (using
heuristic mode analysis arguments) and numerically (using the numerical infrastructure recently
introduced by us).
\end{abstract}
%%%%%%%%%%%%%%%%%%%%%%%%%%%%%%%%%%%%%%%%%%%%%%%%%%%%%%%%%%%%%%%%%%%%%%%%%%%%%%%%%%%%%%%%%%%%%%%%%%%%%%%%%%%%%%%%%%%%%%%%%%%%%%%%%

\section{Introduction}
\label{sec:intro}

The \textit{Nariai spacetime} was discovered by Nariai in 1950 (see the reprints of the original works in \cite{Nariai:1999is,Nariai:1999jz}). 
It is the solution of Einstein's vacuum equations\footnote{We use the signature convention $(-,+,+,+)$ for spacetime metrics and the sign convention for curvature tensors in \cite{Wald:1984un}. In this convention the de-Sitter spacetime is a solution of \Eqref{eq:vEFEc} with $\Lambda>0$.} 
\begin{equation}
  \label{eq:vEFEc}
  G_{ab}+\Lambda g_{ab}=0,
\end{equation}
where\footnote{In this whole paper, we consider $\Lambda$ either as \emph{any} positive constant or set $\Lambda=1$.} $\Lambda>0$ is the cosmological constant and $G_{ab}$ is the Einstein tensor associated with the metric $g_{ab}$,
with spatial topology $\So\times\St$ and
\begin{equation}\label{eq:Nariai}
g_{ab}= \dfrac{1}{\Lambda} (-\df t^{2} +  \cosh^{2} t \: \df  \rho^2 + g_{\mathbb{S}^2} ).
\end{equation}
Here, $\rho$ is the standard coordinate along the spatial $\mathbb{S}^1$-factor and $g_{\mathbb{S}^2}$ is the metric of the standard round unit two-sphere. The time coordinate is $t\in (-\infty,\infty)$. 
In the last years, the Nariai spacetime has become an object of special interest since Ginsparg and Perry \cite{ginsparg1983semiclassical} found that it  can be interpreted as a de-Sitter universe containing a black hole of ``maximal size''. Thanks to its geometrical properties, the Nariai spacetime has been used to model several situations. One of the most remarkable applications was carried out by Bousso and Hawking \cite{bousso1995probability,bousso1996pair,bousso1998anti, Bousso:2003wa} who used this spacetime to study the quantum pair creation of black holes during inflation. These cosmological models, at the borderline between inflation and gravitational collapse, restricted to  spherically symmetric perturbations of the Nariai spacetime. It was found that under certain conditions those models asymptotically approach the de Sitter universe in agreement with the \keyword{cosmic no-hair conjecture} \cite{gibbons77,Hawking:1982ir}. 
It states that this behavior is \emph{generic} for inhomogeneous and anisotropic expanding solutions of \Eqref{eq:vEFEc}. Although there is some mathematical evidence \cite{Kitada:1993wm,Ringstrom:2008kx,Wald:1983if,Andreasson:2016cg} that supports the validity of  this conjecture, the general case still remains unclear.
A particular property of the Nariai spacetime itself is its peculiar time dependence which is \textit{not} consistent with this. While the spatial $\mathbb{S}^1$-factor  expands exponentially for large $t$, the geometry of the spatial $\mathbb{S}^2$-factor remains constant. Thus, the expansion of this solution is very anisotropic, and, in fact, is \textit{inconsistent} with the {cosmic no-hair paradigm}.
In more geometric terms, this corresponds to the fact, which was proven for the first time in \cite{Beyer:2009vm} (an alternative proof was given in \cite{Friedrich:2015tu}), that the Nariai spacetime does not possess even a piece of a   smooth conformal boundary. 
If the cosmic no-hair conjecture really holds, the Nariai spacetime must be therefore ``very special'', and hence in particular be  unstable under ``generic'' perturbations. 

In \cite{Beyer:2009vm} one of us has initiated the study of \textit{homogeneous} (but fully nonlinear) perturbations of the Nariai spacetime.  The Nariai solution is a member of the class of \keyword{Kantowski-Sachs} spatially homogeneous (but anisotropic) solutions \cite{Wainwright:2005wss} of Einstein's vacuum equation with a positive cosmological constant. The spatial topology of all of these models is $\So\times\St$. It was found there that the Nariai solution is \keyword{critical} in this family in the following sense. For all Kantowski-Sachs models, except for the Nariai solution, we can choose the time orientation such that the solution either collapses to the future (a \keyword{big crunch} characterized by the formation of a curvature singularity and all future inextendible causal curves are incomplete) or expands eternally to the future in consistency with the {cosmic no-hair picture}  (existence of a smooth future conformal boundary in consistency with the future asymptotics of de-Sitter space so that all future inextendible causal curves are complete). 
The Nariai solution is exactly at the borderline between these two extremes  as the curvature is bounded everywhere and all inextendible causal curves are both future and past complete, but it nevertheless does not agree with cosmic no-hair.
 
The first rigorous work in \cite{Beyer:2009vm} on this topic has recently been extended in \cite{Fajman:2016vf}. The numerical studies in \cite{Beyer:2009ta} of Gowdy-symmetric \cite{Gowdy:1974hv,Chrusciel:1990ti}  (see \Sectionref{sec:formulationeqs} for more details on Gowdy symmetry) inhomogeneous fully nonlinear perturbations of the Nariai solution have revealed 
evidence that the analogous critical phenomenon also exists in much larger classes of spacetimes. In particular, it was found that all solutions, which are obtained from initial data not too far away from the Nariai solutions, always either globally collapse or expand in the same manner as in the spatially homogeneous case --- with the exception of \emph{critical solutions} which are exactly at the borderline between these two cases.
In particular, it was found that in contrast to the spherically symmetric models considered by Bousso et al.\ above \cite{Bousso:2003wa}, Gowdy symmetric models never \emph{locally} collapse or expand,  and  the formation of cosmological black holes in this class was therefore ruled out. Because the perturbations considered in \cite{Beyer:2009vm} were small in some sense, the question remained open  whether this may be different for larger Gowdy symmetric perturbations.  One of the finding in our paper here now suggests that Gowdy symmetric models indeed never form cosmological black holes. For future work, it will be interesting to pose this question again within more general classes of spacetimes, for example $U(1)$-symmetric spacetimes, and study whether cosmological black holes may be created by perturbations of the Nariai spacetime. 

Before we continue, let us remind the reader about the heuristic idea of the study of the criticality of the cosmological models in \cite{Beyer:2009ta}. There we worked with Gowdy symmetric initial data (which satisfy the  constraint equation implied by \Eqref{eq:vEFEc}) given by two real parameters $\mu$ and $\Sigma_\times^{(1)}$ whose precise definition is irrelevant now (cf.\ \cite{Beyer:2009ta} for the details). The special choice $\mu=\Sigma_\times^{(1)}=0$ corresponds to Nariai initial data while $\Sigma_\times^{(1)}=0$ and $\mu\in\R$ yields a class of spatially homogeneous models. The larger the value of $|\Sigma_\times^{(1)}|$ is, however, the ``more spatially inhomogeneous'' the initial data are. The idea was to fix some non-zero value of the ``inhomogeneity parameter''  $\Sigma_\times^{(1)}$ and then to study a sequence of (fully nonlinear) cosmological models given by a sequence of values of $\mu$. On the one hand, it was found that if $\mu$ is sufficiently large, the corresponding model expands globally to the future; in fact, the solution develops a smooth conformal boundary to the future in this case and is hence fully consistent with the cosmic no-hair conjecture\footnote{Notice that the studies in \cite{Beyer:2009ta} made use of Friedrich's conformal field equations \cite{Friedrich:2004vx} and therefore allowed us to calculate the \textit{full} conformally compactified solutions, including the conformal boundary if it exists. Here, we will not make use of such compactification techniques.}. If $\mu$ is sufficiently small on the other hand, the model collapses globally to the future and eventually forms a curvature singularity. At the borderline between these two regimes 
corresponding to a critical value for $\mu$, the corresponding model \textit{neither} collapses nor expands to the future. However, no further information about such critical models was extracted in \cite{Beyer:2009ta}.

The purpose of our present paper is manifold. Again, we restrict to the class of Gowdy-symmetric models with a positive cosmological constant and we revisit the same situation, but tackle it with a more advanced approach. To this end, we use a different class of initial data which now depends on \textit{three} parameters $\epsilon$ (which has a similar meaning as the ``inhomogeneity parameter'' $\Sigma_\times^{(1)}$ above), $C$ (which has a similar meaning as the parameter $\mu$ above) and an additional parameter $\ell$ which essentially controls the wave number of the inhomogeneous perturbation (the initial data in \cite{Beyer:2009ta} restricted to the case $\ell=2$). The details are discussed in \Sectionref{sec:familyIDN}.
On the one hand, we  confirm and strengthen the findings in \cite{Beyer:2009ta} by performing a similar numerical analysis. On the other hand, however, we shall focus  in great detail on the \textit{critical solutions} here and thereby reveal an interesting new oscillatory phenomenon which could potentially be interpreted as gravitational waves. The main finding of our paper are now summarized as three main results. The purpose of this paper is to provide the details and give justifications.

In all of what follows we shall assume\footnote{In this paper, we shall not be interested in the homogeneous perturbations associated with the case $\ell=0$. The case $\ell=1$ is a special borderline case which turns out to be not well described by our analytic method discussed in \Sectionref{sec:heuristicmode}. We therefore completely disregard the case $\ell=1$ in this paper.} $\ell\ge 2$. 
Notice that the well-understood \cite{Beyer:2009vm} homogeneous case of our models corresponds to $\epsilon=0$. The Nariai solution is determined by $\epsilon=C=0$. One can easily check that if $\epsilon\not=0$ or $C\not=0$,  the corresponding solution of \Eqref{eq:vEFEc} is not isometric to the Nariai solution by comparing the Kretschmann scalar with the particular globally constant value for the Nariai solution. Our first main finding is summarized as follows.
\begin{result}
  \label{conj1}
  Pick any real value $\epsilon$ and integer $\ell\ge 2$. Then there is a constant $C_{crit}$ such that the solution of \Eqref{eq:vEFEc}, determined by initial data given by the parameter $\epsilon$, $\ell$ and any real value $C$ as in \Sectionref{sec:familyIDN},
globally collapses and forms a curvature singularity if $C>C_{crit}$ and globally expands in consistency with the cosmic no-hair conjecture if $C<C_{crit}$.
\end{result}

For small values of $\epsilon$, this result is in full consistency with the findings in \cite{Beyer:2009ta}.
Here we claim now that this also holds for large values of $\epsilon$. As mentioned earlier, this rules out in particular the possibility of ``local'' collapse and hence there are generically indeed only two kinds cosmological models in this class. In this paper here we provide some  more refined numerical evidence complemented by a heuristic mode analysis (see \Sectionref{sec:heuristicmode}).
We call any of our models \emph{critical} if $C=C_{crit}$ for any given $\epsilon$ and $\ell$, and \emph{almost critical} or \emph{close-to-critical} if $C\not= C_{crit}$, but $|C-C_{crit}|\ll 1$.

The second main finding of our work which significantly goes beyond the results in \cite{Beyer:2009ta}  is summarized as follows.
  \begin{result}
    \label{conj2}
    For any non-zero value of $\epsilon$, the critical and close-to-critical solutions asserted in Result~\ref{conj1} are oscillatory.
  \end{result}
Based on the before-mentioned heuristic analysis in \Sectionref{sec:heuristicmode}, we are able to derive formulas for oscillation frequencies, amplitudes and phases and how these depend on the initial data  parameters. 
The only non-oscillatory solutions correspond to the spatially homogeneous case $\epsilon=0$ in which the critical solution is known to coincide with the Nariai solution (see \cite{Beyer:2009vm}) and therefore has the peculiar asymptotics discussed above. Our numerical work here suggests that \emph{all} the critical models, also the inhomogeneous ones, are \keyword{Nariai-like} in the following sense.
\begin{result}
  \label{conj3}
  The critical solutions behave asymptotically as follows. While the spatial \So-factor expands exponentially,  the spatial \St-factor geometry oscillates around the round unit $2$-sphere geometry and is therefore in particular bounded. All these models therefore violate the cosmic no-hair picture by these highly anisotropic asymptotics.
\end{result}

We emphasize that our work here is \textit{not} actually concerned with the \textit{instability of the Nariai solution}; this issue is addressed elsewhere \cite{Beyer:2009vm, Fajman:2016vf}. The point of our work here is now to identify and describe {inhomogeneous} critical models and their Nariai-like asymptotics \textit{all} of which violate the cosmic no-hair paradigm.
We remark that all our numerical studies were conducted with slight generalizations of the numerical code presented in \cite{Beyer:2016fc}. More details and references regarding our numerical infrastructure are given in \Sectionref{sec:evolutionN}.

The paper is organized as follows. In \Sectionref{sec:setup}, we discuss the general setup, i.e., the formulation of Einstein's equation in the presence of symmetries via \keyword{Geroch's symmetry reduction}, our extraction of evolution equations with a well-posed initial value problem and of constraint equations from this, our numerical implementation and our particular family of initial data. \Sectionref{sec:criticalNariai} is devoted to our analytical and numerical studies. First we discuss our heuristic mode analysis which is the basis for all of what follows. Then we provide numerical evidence for all of the main results above.

%%%%%%%%%%%%%%%%%%%%%%%%%%%%%%%%%%%%%%%%%%%%%%%%%%%%%%%%%%%%%%%%%%%%%%%%%%%%%%%%%%%%%%%%%%%%%%%%%%%%%%%%%%%%%%%%%%%%%%%%%%%%%%%%%

\section{Setup: Formulation and implementation of Einstein's equations}
\label{sec:setup}

\subsection{Einstein's vacuum equations for Gowdy symmetry and spatial \texorpdfstring{$\So\times\St$}{S1xS2}-topology}
\label{sec:formulationeqs}

\paragraph{Geroch's symmetry reduction} As mentioned earlier, we shall focus here on Gowdy symmetric spacetimes with spatial topology $\So\times\St$. We start by equipping the spatial manifold $\So\times\St$ with coordinates $(\rho,\theta,\varphi)$ where $\rho\in (0,2\pi]$ is the standard parameter on the $\So$-factor and $(\theta,\varphi)$ are standard polar coordinates on the $\St$-factor. With respect to these coordinates it turns out that a spacetime with spatial $\So\times\St$-topology is Gowdy (or $\U\times\U$-)symmetric if the metric is invariant both under translations along the $\So$-factor (i.e., is independent of $\rho$), and, under rotations of the $\St$-factor around the polar coordinate axis (i.e., is independent of $\varphi$).  One can show that Einstein's equations for this class of spacetimes can be formulated in almost exactly the same way as for the class of Gowdy symmetric spacetimes with spatial $\Sth$-topology which was discussed in detail in \cite{Beyer:2016fc}. Because of the close similarity we shall refer to that paper for all the details and only give a quick summary of the necessary results now.

When Geroch's symmetry reduction \cite{Geroch:1971ix} is performed with respect to the Gowdy Killing vector fields $\xi^a=\partial_\rho^a$ for any $3+1$-dimensional Gowdy-symmetric metric $g_{ab}$ 
in a spacetime $M=\R\times\So\times\St$, one finds that Einstein's vacuum field equations with cosmological constant \eqref{eq:vEFEc} imply the system
\begin{eqnarray}\label{eq:geroch_einstein_equations}
 \nabla_{a} \nabla^{a} \psi   &=& \dfrac{1}{\psi} \left(  \nabla_{a} \psi \nabla^{a} \psi  -  \nabla_{a} \omega \nabla^{a} \omega \right) - 2 \Lambda, \nonumber\\
 \nabla_{a} \nabla^{a} \omega &=& \dfrac{2}{\psi} \nabla^{a} \psi \nabla_{a} \omega , \\
R _{ab} &=&   \dfrac{1}{2\psi^2 } \left(  \nabla_{a} \psi \nabla_{b} \psi + \nabla_{a} \omega \nabla_{b} \omega \right) + \dfrac{2\Lambda}{\psi} h_{ ab }   , \nonumber 
\end{eqnarray}
on the $2+1$-manifold $S=\R\times\St$ where $h_{ab}$ is a metric on $S$ with signature $(-,+,+)$, $\nabla_a$ is its covariant derivative and $R_{ab}$ is the corresponding Ricci tensor. The scalar field $\psi$ is defined on $M$ as
\begin{equation}
\psi := g_{ab}\xi^a\xi^b,
\end{equation}
for $\xi^a=\partial_\rho^a$ which is then projected\footnote{In order to simplify the notation of \cite{Beyer:2016fc}  slightly, we shall not distinguish here between quantities on $S$ and their counterparts on $M$ which are obtained by a pullback along the projection map $\pi$. In fact, all quantities, which carry a $\tilde{\,\,}$ in \cite{Beyer:2016fc}, shall be written without a $\tilde{\,\,}$ here.} to $S$. 
The other scalar field $\omega$ is the well-defined global potential of the twist $1$-form $\Omega_a$ on $M$ defined by
\begin{equation}
\nabla_a\omega=\Omega_a := \epsilon_{abcd} \xi^{b} \mathfrak D^c \xi^{d}.
\end{equation}
Here, $\mathfrak D_c$ is any derivative operator (for instance the covariant derivative associated with $g_{ab}$) and $\epsilon_{abcd}$ is a volume form associated with $g_{ab}$.
We shall often refer to $\psi$ as the \textit{norm} and to $\omega$ as the \textit{twist} of $\xi^a$ respectively. 
\Eqsref{eq:geroch_einstein_equations} can be interpreted as the $2+1$-dimensional Einstein equations\footnote{Strictly speaking, this is only the case when $\Lambda=0$. When $\Lambda\not=0$, the second term on the right-hand side of the third equations differs from a standard cosmological constant term. This will however not play any role in our discussions here.} coupled to two scalar fields $\psi$ and $\omega$. 

Without going into the details, see for example \cite{Beyer:2016fc}, let us mention that once a solution $(h_{ab}, \psi, \omega)$ of \Eqsref{eq:geroch_einstein_equations} has been found on $S$, one can construct the corresponding physical spacetime metric $g_{ab}$ on $M$ which then solves \Eqref{eq:vEFEc}. It is important to notice that $g_{ab}$ and $h_{ab}$ are related as follows
\begin{equation}\label{ec:21metric}
\hat{h}_{ab} := g_{ab} − \frac{1}{{\psi}} \xi_a \xi_{b},
\end{equation}
and
\begin{equation}\label{eq:conformal_metric}
h_{ab} := \psi \hat{h}_{ab}.
\end{equation}
The metric $h_{ab}$ in \Eqsref{eq:geroch_einstein_equations} is therefore not the \textit{physical} $2+1$-metric, but is related by the conformal transformation \Eqref{eq:conformal_metric} to the physical $2+1$-metric $\hat{h}_{ab}$ defined by \Eqref{ec:21metric}.
Both metrics $h_{ab}$ and $\hat{h}_{ab}$ will play a role in our discussion later.

The system of equations \eqref{eq:geroch_einstein_equations} takes care of only \textit{one} of the two Gowdy Killing vector fields $\xi^a=\partial_\rho^a$ (i.e., translations along the spatial $\So$-factor) so far. It turns out that the second Killing field $\partial_\varphi^a$ prevails on $S$, i.e., all quantities  $h_{ab}$, $\psi$ and $\omega$ on $S$ defined above are \keyword{axi-symmetric} and therefore invariant under the action of $\partial_\varphi^a$ (i.e., under rotations around the polar coordinate axis of $\St$). We remark that one could  perform Geroch's symmetry reduction a second time, but now with respect to this Killing field (see also \cite{Geroch:1972bp}). This however leads to explicit singularities at the poles of the two-sphere. As in \cite{Beyer:2016fc}, we shall therefore work in all of what follows with axially symmetric (i.e., $\varphi$-invariant) solutions of \Eqsref{eq:geroch_einstein_equations} without any further symmetry reductions.

\paragraph{The (generalized) wave map formalism}
 The next task is to extract suitable evolution and constraint equations from \Eqsref{eq:geroch_einstein_equations} in order to obtain a well-posed initial value problem.  The first two equations of \eqref{eq:geroch_einstein_equations} are scalar wave equations. It therefore remains to deal with the third equation of \eqref{eq:geroch_einstein_equations}. Since this is the $2+1$-Einstein equation for the metric $h_{ab}$ with a (as one can check) divergence free energy momentum tensor of the ``matter source'', we can apply all kinds of standard techniques which were developed for $3+1$-Einstein's equations. Because of its geometric nature, which is particularly useful for dealing with the spatial topology $\St$ of $S$, we work with the generalized wave map formalism \cite{Friedrich:1991nn} here. Again, all details are worked out in \cite{Beyer:2016fc} and we just give a quick summary here.

The point is that the third equation in \Eqsref{eq:geroch_einstein_equations} is a-priori not a system of wave equations for the components of $h_{ab}$ (with respect to any  frame) and hence the initial value problem is in general not well-posed. This problem is overcome  when we replace $R_{ab}$ in that equation by the new tensor field
\begin{equation}
\label{eq:riccihat}
\hat R_{ab}:= R_{ab}+\nabla_{(a} \mathcal{D}_{b)},
\end{equation}
where the components $\mathcal{D}^\alpha$ of the vector field $\mathcal{D}^{b}$ with respect to any smooth local frame\footnote{The components of tensor fields with respect to any such frame on $S=\R\times\St$ are denoted by greek indices.} are given as
\begin{equation}
\label{eq:defDs}
\mathcal{D}^\alpha := 
(-\Gamma^\alpha{}_{\beta\gamma}  + \bar\Gamma^\alpha{}_{\beta\gamma})
h^{\beta\gamma} + f^\alpha.
\end{equation}
The vector field $f^a$ can be specified freely and is referred to as a \keyword{gauge source field}. Its components $f^\alpha$ with respect to any frame are often called \keyword{gauge source functions}. The connection coefficients of the covariant derivative associated with $h_{ab}$ with respect to this frame are denoted above by $\Gamma^\alpha{}_{\beta\gamma}$, while $\bar\Gamma^\alpha{}_{\beta\gamma}$ are the corresponding connection coefficients associated with any freely specifiable  \keyword{reference metric} $\bar h_{ab}$ on $S$.
In total this produces a (complicated) system of quasilinear wave equations for the components of the metric and the fields $\psi$ and $\omega$:
\begin{eqnarray*}
 \nabla_{a} \nabla^{a} \psi   &=& \dfrac{1}{\psi} \left(  \nabla_{a} \psi \nabla^{a} \psi  -  \nabla_{a} \omega \nabla^{a} \omega \right) - 2 \Lambda, \nonumber\\
 \nabla_{a} \nabla^{a} \omega &=& \dfrac{2}{\psi} \nabla^{a} \psi \nabla_{a} \omega , \\
\hat R _{ab} &=&   \dfrac{1}{2\psi^2 } \left(  \nabla_{a} \psi \nabla_{b} \psi + \nabla_{a} \omega \nabla_{b} \omega \right) + \dfrac{2\Lambda}{\psi} h_{ ab }. \nonumber 
\end{eqnarray*}
In particular, the initial value problem of these \keyword{evolution equations}
is well-posed for suitable initial data.

Suppose now that a solution $(h_{ab},\psi,\omega)$ of the initial value problem of the evolution equations has been found on $S$. It is clear that this is a solution of the original system \eqref{eq:geroch_einstein_equations} if $\mathcal{D}^a$ vanishes and hence $\hat R_{ab}= R_{ab}$ on $S$. In this case we say that $h_{ab}$ is in \keyword{generalized wave map gauge}. We show now that $\mathcal{D}^a$ vanishes only if the initial data for the evolution equations satisfies certain constraints. As discussed for example in \cite{Ringstrom:2009cj},
the evolution equations, the fact that the energy momentum tensor of the matter source in \Eqsref{eq:geroch_einstein_equations} is divergence free, and the contracted Bianchi identities together imply
\begin{equation}\label{subsidiarysystem}
\nabla_{b} \nabla^{b} \mathcal{D}_{a} - \mathcal{D}^{b}  R_{ab}= 0.
\end{equation}
Since the metric $h_{ab}$ (and hence $R_{ab}$) is considered as known at this stage, this is a linear homogeneous system of wave equations for the unknown $\mathcal{D}_a$. It follows that $\mathcal{D}^a$ vanishes everywhere on $S$ if and only if $\mathcal{D}^a=0$ and $\nabla_a \mathcal{D}^b=0$ on the initial hypersurface; these two conditions therefore constitute \textit{constraints}.
The first constraint takes the form (with respect to any smooth local frame)
\begin{equation}
\label{eq:constraint1}
  0=\mathcal{D}^\nu=h^{\rho\sigma}(\bar\Gamma^{\nu}{}_{\rho\sigma}-\Gamma^{\nu}{}_{\rho\sigma})+f^\nu,
\end{equation}
which can be satisfied for \emph{any} initial data $h_{ab}$, $\psi$ and $\omega$ on the initial hypersurface by a suitable choice of the free gauge source quantities $f^a$ and $\bar h_{ab}$. \Eqref{eq:constraint1} is therefore referred to as the \textit{gauge constraint}. Once we know that the gauge constraint is satisfied, it turns out that the second constraint above
\begin{equation}
  \label{eq:constraint2}
  \nabla_\mu \mathcal{D}_\nu=0
\end{equation}
is equivalent to the standard Hamiltonian and Momentum constraints which we discuss in more detail in \Sectionref{sec:familyIDN}. We emphasize the surprising fact that \Eqref{eq:constraint2} is \emph{not} another restriction on the gauge source quantities $f^a$ and $\bar h_{ab}$; these are only constrained by the gauge constraint.  \Eqref{eq:constraint2} therefore represents the ``physical constraint'' imposed on the initial data for $h_{ab}$, $\psi$ and $\omega$.

\subsection{Formulation and implementation of the evolution equations}
\label{sec:evolutionN}

\paragraph{Choice of frame and parametrization of the metric}
If $t$ is any time function on $S$ and $(t,\theta,\varphi)$ are coordinates as before, 
we set 
\begin{equation}\label{eq:referenceframe}
 T^a:=\partial^a_t, \quad m^a:=\frac 1{\sqrt 2}\left(\partial^a_{\theta}-\frac i{\sin\theta}\partial^a_\varphi\right).
\end{equation}
Then we choose
\begin{equation}
  \label{eq:referenceframe3}
  (\partial_0^a,\partial_1^a,\partial_2^a)=(T^a,m^a,\mbar^a)
\end{equation}
as our local frame which is defined almost everywhere on $S=\R\times \St$ (excluding the poles of the two-sphere). 
Notice, that this frame is in general not an orthonormal frame with respect to $h_{ab}$. It is merely a particular linear combination of the coordinate frame which is motivated  by the spin-weight formalism below.
In the following we shall express all tensor fields on $\St$ with respect to this frame and its dual frame  $( \omega^0_a ,\omega^1_a, \omega^2_a )$ which is given by
\begin{equation}
\label{eq:referenceframe2}
\omega_a^0 = \nabla_a t , \quad  \omega^1_a = \frac 1{\sqrt 2} \left( \nabla_a \theta+i \sin\theta \nabla_a \varphi \right), \quad \omega^2_a = \overline{\omega}^1_a.
\end{equation}
It terms of this frame, we can write
\begin{equation}\label{3Dmetric_smoothframe}
h_{ab} =   \lambda \hspace{0.1cm} \omega_a^0 \omega_b^0  + 2 \hspace{0.1cm} \omega_{(a}^0 \left(  \beta \hspace{0.1cm} \omega_{b)}^1 + \bar{\beta} \hspace{0.1cm} \omega_{b)}^2  \right)  +  2 \delta \hspace{0.1cm} \omega_{(a}^1 \omega_{b)}^2  + \phi \hspace{0.1cm} \omega_a^1 \omega_b^1 + \bar{ \phi } \hspace{0.1cm} \omega_a^2 \omega_b^2.
\end{equation}

For the spin-weight formalism \cite{Penrose:1984tf,Beyer:2016fc,Beyer:2014bu,Beyer:2015bv,Beyer:2012ti} we assume that the fields $T^a$, $m^a$ and $\mbar^a$ have spin-weights $0$, $+1$ and $-1$, respectively, which implies that the spin-weights of $\omega^0_a$, $\omega^1_a$, $\omega^2_a$ are $0$, $-1$ and $+1$, respectively. The quantities $\lambda,\delta,\beta$ and $\phi$ in \Eqref{3Dmetric_smoothframe} therefore have spin-weights $0$, $0$, $+1$, $+2$, respectively, and the complex conjugates carry the corresponding negative spin-weights.
It is of fundamental importance for all of what follows that once the gauge freedom in terms of the smooth quantities $f^a$ and $\bar h_{ab}$ has been fixed, the whole system of evolution equations can be written as a quasilinear coupled system of six complex wave equations for the six complex unknowns $\lambda$, $\delta$, $\beta$, $\phi$, $\psi$ and $\omega$. Moreover, once all directional derivatives along $m^a$ and $\bar m^a$ have been replaced by the so-called $\eth$- and $\bar\eth$-operators via \Eqref{eq:ethm}, each term in each equation of this system has a consistent well-defined spin-weight and is \emph{explicitly regular at the poles} $\theta=0$ and $\theta=\pi$ of the $2$-sphere. Indeed, this explicit regularization of the ``pole problem'' is the main advantage of the spin-weight formalism.

We recall that Gowdy symmetry implies that $\partial_\varphi^a$ is a Killing vector field on $S$. Since $\partial_\varphi^a$ commutes with each of the fields in \Eqsref{eq:referenceframe} and \eqref{eq:referenceframe2}, it follows that $h_{ab}$, as given by \Eqref{3Dmetric_smoothframe}, is invariant under the action of $\partial_\varphi^a$ if and only if all the quantities $\lambda,\delta,\beta$ and $\phi$ are functions of $t$ and $\theta$ only. This means in particular that all these functions can be expanded in terms of  axi-symmetric spin-weighted spherical harmonics, see \Sectionref{Sec:swsh}.
We also know that all quantities with spin-weight $0$ must be real, while other quantities could in principle be complex. However, in the particular representation in $(\theta,\varphi)$-coordinates used exclusively in this whole paper, one can show  that if \emph{all} unknown quantities (and their time derivatives) in the evolution equations are real  at the initial time, if all gauge source quantities 
\[f_0=f_a T^a,\quad f_1=f_a m^a,\quad f_2=f_a\bar m^a\]
are real for all times, and, if
 the background metric $\bar h_{ab}$ is once and for all chosen as
\begin{equation}\label{eq:background_metric}
  \bar{h}_{ab} = -\omega_a^0 \omega_b^0  +  2\omega_{(a}^1\omega_{b)}^2
\end{equation}
for all times,
 then all unknown quantities are real for all times. Below we see that this restriction to real quantities is purely a gauge restriction. In summary, without further notice we shall assume in all of what follows that all quantities 
$\lambda$, $\delta$, $\beta$, $\phi$, $\psi$, $\omega$, $f_0$, $f_1=f_2$  are real and only depend on $t$ and $\theta$.

\paragraph{Gauge drivers and conformal time gauge}
Instead of fixing the gauge freedom by choosing the gauge source quantities $f_0$ and $f_1=f_2$ as outlined in the previous section, it may sometimes be advantageous numerically to fix the gauge by choosing 
the ``lapse'' and ``shift'' of the $2+1$-dimensional metric $h_{ab}$. 
The equations which then determine $f_0$ and $f_1$ are often called \keyword{gauge drivers}. Even though this approach has been used successfully in some situations (see for instance \cite{Preto2005:Evolution-of-Binary-Black-Hole,pretorius2005numerical}), it may cause numerical instabilities. The reason lies in the fact that the resulting total system of evolution equations (including the gauge drivers) may not have a well-posed initial value problem despite the fact that the original evolution equations in the wave gauge formalism do. 
Some general proposals for gauge drivers which do not suffer from this problem  can be found in \cite{Lindblom:2008kk,Lindblom:2009in}. In this work here now, we will construct particular gauge drivers now and then show that the total system of evolution equations is strongly hyperbolic.

To start with, we consider the unit normal vector to the $t=const$-surfaces (recall that $\beta$ is assumed to be real)
\begin{equation*}\label{normal_vector}
n^{a}= \frac{1}{\alpha}\left(T^a-\beta (m^a+\mbar^a)\right), \quad  n_{a} =  -\alpha\omega_a^0;
\end{equation*}
recall \Eqsref{eq:referenceframe}, \eqref{eq:referenceframe2} and \eqref{3Dmetric_smoothframe},
where
\[\alpha = \sqrt{ \beta^2 - \lambda  }. \]
We can therefore interpret $\beta$ as the \keyword{shift}\footnote{The shift vector is $\beta^a=\beta (m^a+\mbar^a)$.} and $\alpha$
as
the \keyword{lapse}. The induced metric on the $t=\mathrm{const}$-hypersurfaces is therefore (recall that $\phi$ is assumed to be real)
\begin{equation}
\label{eq:generalinducedmetric}
\gamma_{ab} = h_{ab} + n_{a} n_{b}
=\beta^2 \hspace{0.1cm} \omega_a^0 \omega_b^0  + 2\beta \hspace{0.1cm} \omega_{(a}^0 \left(\omega_{b)}^1 + \omega_{b)}^2  \right)  +  2 \delta \hspace{0.1cm} \omega_{(a}^1 \omega_{b)}^2  + \phi \left(\omega_a^1 \omega_b^1 + \omega_a^2 \omega_b^2\right),
\end{equation}
cf.\ \Eqref{3Dmetric_smoothframe}.
Now, according to our discussion of gauge drivers above, let us attempt to fix the gauge by picking 
\begin{equation}
\alpha=\delta,\quad \beta=0,
\end{equation}
during the whole evolution.
This corresponds to
\begin{equation}
\label{eq:conformaltimegauge}
 \lambda = -\delta , \quad \beta = 0.	
 \end{equation} 
Heuristically, the idea is that the lapse is proportional to the area $\delta$ of the spatial $2$-sphere. From the point of view of any Eulerian observer, the coordinate clock will therefore tick faster or slower depending on whether the $2+1$-spacetime is expanding or collapsing. An important consequence is that the foliation tends to ``freeze'' in the collapsing case. This gauge therefore avoids singularities. This sort of gauge  is commonly known in the standard cosmology literature as \keyword{conformal time gauge}, and is used frequently in the linear theory of cosmological perturbations \cite{Mukhanov:1992vf}. 

In this work, we wish to implement gauge drivers which preserve this gauge during the evolution. To do so, we use  \Eqref{eq:conformaltimegauge} to express the evolution equations for $\lambda$ and $\beta$ in the wave gauge formalism as evolution equations  for the gauge source functions $f_0$ and $f_1$. Hence, from now on, $\lambda$ and $\beta$ will not be considered as unknown variables anymore, but $f_0$ and $f_1$ will. The question is whether the resulting evolution system is hyperbolic, and, if yes, in which sense. In what follows we consider this question in detail.

We continue to assume that all unknown fields in the evolution equations are real functions,  and all the partial derivatives with respect to the coordinate $\varphi$ vanish. Then, expanding the covariant derivatives and expressing the frame vectors $m^a$ and $\overline{m}^a$ in terms of the coordinate vector $\partial_{\theta}$,  we obtain  evolution equations for $\delta$, $\phi$, $\psi$ and $\omega$ of the form
\begin{eqnarray}
\partial_{tt} \delta + a \partial_{\theta \theta} \delta  + b\partial_{\theta} f_{1} = ... \:,\label{ec:ec:second_ordersystem1} \\
\partial_{tt} \phi   + a \partial_{\theta \theta} \phi + b\partial_{\theta} f_{1} = ... \:,\label{ec:ec:second_ordersystem2}  \\
\partial_{tt} \psi   + a \partial_{\theta \theta} \psi  = ... \:,\label{ec:ec:second_ordersystem3} \\
\partial_{tt} \omega + a \partial_{\theta \theta} \omega  = ... \:,\label{ec:ec:second_ordersystem4} 
\end{eqnarray}
where $a = h^{11}/h^{00} $ and $b= \sqrt{2}/h^{00} $. Note that we have used $f_2=f_1$ in the evolution equation for $\delta$. The ellipses in the right-hand side of the equations denote lower order terms which are irrelevant for this analysis. Setting $\lambda = -\delta$ and $\beta = 0$ we obtain  evolution equations for the gauge source functions $f_0$ and $f_1$, respectively  as
\begin{eqnarray}
 \partial_{tt} \lambda  = -\partial_{tt} \delta & \Longrightarrow & \partial_{t} f_0 - \partial_{\theta} f_{1} = ... \ ,\label{ec:ec:first_ordersystemf1} \\
 \partial_{tt} \beta   =  0            & \Longrightarrow & \partial_{t} f_1 - \partial_{\theta} f_{0} = ... \ . \label{ec:ec:first_ordersystemf2}
\end{eqnarray}
Naturally, these evolution equations, which we call \textit{gauge drivers}, control the behavior of the  generalized gauge source functions such that the conformal time gauge is preserved during the evolution. Next, in order to analyze the hyperbolicity of the resulting system of evolution equations \Eqsref{ec:ec:second_ordersystem1}--\eqref{ec:ec:first_ordersystemf2},  we rewrite it in first-order form as
\begin{equation}\label{firs_order_system_of_equations0}
\partial_{t} u + \Pi \ \partial_{\theta} u = s(u),	
\end{equation}
where we have defined the vector 
\begin{eqnarray*}\label{firs_order_variables}
u = \left(  \delta , \partial_{t} \delta , \partial_{\theta} \delta, \phi, \partial_{t} \phi , \partial_{\theta} \phi, \psi, \partial_{t} \psi , \partial_{\theta} \psi , \omega, \partial_{t} \omega , \partial_{\theta} \omega , f_0 ,  f_1   \right),
\end{eqnarray*}
and where $\Pi$ is a $14 \times 14$ (non-symmetric) matrix\footnote{Because of the size of $\Pi$ we do not write it here explicitly.}  with eigenvalues 
\begin{eqnarray*}
\upsilon_{1,2,3,4} = - \upsilon_{5,6,7,8} =  -\sqrt{-a}, \quad \upsilon_{13}  =  - \upsilon_{14} = -1, \quad \upsilon_{9,10,11,12} =0. \nonumber
\end{eqnarray*}
After a straightforward calculation we show that all eigenvectors are linearly independent. 
Note that $h^{00},-h^{11}>0$ provided that $h_{ab}$ is a Lorentzian metric with signature $(-,+,+)$, thus $a<0$ and hence all the eigenvalues of $\Pi$ are real. This implies that $\Pi$ has a complete set of eigenvectors with real eigenvalues. Our system \Eqref{firs_order_system_of_equations0} is therefore \keyword{strongly hyperbolic}. This implies the well-posedness of the initial value problem.

\paragraph{Constraint damping terms}
In order to deal with the widely known  problem of the growth of constraint violations during numerical evolutions,  Brodbeck et al.\ \cite{Brodbeck:1999en} have suggested a general approach 
such that the constraint surface is an attractor. Later, following this idea, Gundlach et al.\ \cite{Gundlach05} introduced so called \textit{constraint damping terms} into Einstein's equations by adding to the right side of \Eqref{eq:riccihat} the term
\begin{equation}\label{constrains_damping_terms}
\kappa \left( \eta_{(a}  \mathcal{D}_{b)}  - g_{a b} \eta^{c} \mathcal{D}_{c} \right),
\end{equation}
with $\eta_{a}$ being a timelike vector and $\kappa$ a constant. With this new term, the subsidiary equation \Eqref{subsidiarysystem} takes the form
\begin{eqnarray}\label{subsydiary_equation2}
\nabla_b \nabla^b \mathcal{D}_{a} - \mathcal{D}^{d}  R_{ad} = 2 \kappa \nabla^{c}  \eta_{(a}  \mathcal{D}_{c)}.\end{eqnarray}
They showed by means of perturbations of the Minkowski spacetime that all the ``short wave length'' modes in the solutions of the subsidiary system \Eqref{subsydiary_equation2} are damped if $\kappa>0$ at either the rate $e^{-\kappa t}$ or $e^{-\kappa t/2}$. In the last years, a good amount of numerical simulations have been successfully conducted  using this approach (see for instance \cite{Lindblom:2009in,Preto2005:Evolution-of-Binary-Black-Hole}), which confirms its effectiveness for several situations. 

Nevertheless, a complete understanding of how the ``long wave length modes'' solutions are damped (or not) for generic spacetimes is still missing. Due to the expanding (or collapsing) behavior of most cosmological spacetimes, the ``long wave length modes''  are expected to be dominant during  the evolution. For our particular interest here it is therefore a-priori not clear whether these constraint damping terms really improve the evolution of constraint violations. 
In order to address this question, we simplify our analysis now by using the assumption that  the  violation covector is approximately only $t$-dependent
\begin{equation}\label{ec:no_time_constraint}
\mathcal{D}_{\mu}(t,\theta) \approx \mathcal{D}_{\mu}(t).
\end{equation}
However, since $\mathcal{D}_\mu$ is a covector, the projection of its spatial components $\mathcal{D}_{1}$ and $\mathcal{D}_{2}$ to the frame $(m^a,\mbar^a)$ must have spin-weight $1$ and $-1$, which directly implies that $\mathcal{D}_{1}=\mathcal{D}_{2}=0$ under the above assumption  because only a function with spin-weight $0$ can have a mode of $l=0$ (which is spatially independent). If the perturbed metrics are ``close'' to the Nariai metric during the initial part of the evolution, we can use it as the background metric for writing the subsidiary equation that rules the evolution of  $\mathcal{D}_{0}$. Thus, replacing the Nariai metric in \Eqref{subsydiary_equation2} with $\eta_a=(-1,0,0)$, we obtain  the evolution equation that rules the behavior of $\mathcal{D}_0(t)$ as
\begin{equation*}
\partial_{tt} \mathcal{D}_{0}(t)  +2 \kappa  \partial_{t}\mathcal{D}_0(t) = 0,
\end{equation*}
for the same constant $\kappa$ as in \Eqref{constrains_damping_terms}. Evidently, the solution of this equation is 
\begin{equation}\label{ec:constrainsolution}
\mathcal{D}_0 (t) = A e^{-2 \kappa t } + B, 
\end{equation}
where $A$ and $B$ are constants. The constraint violation $\mathcal{D}_0(t)$ does therefore not grow if $\kappa > 0$. 

\begin{figure}[t] 
  \centering
   \begin{minipage}{0.49\linewidth} 
   \includegraphics[width=\textwidth]{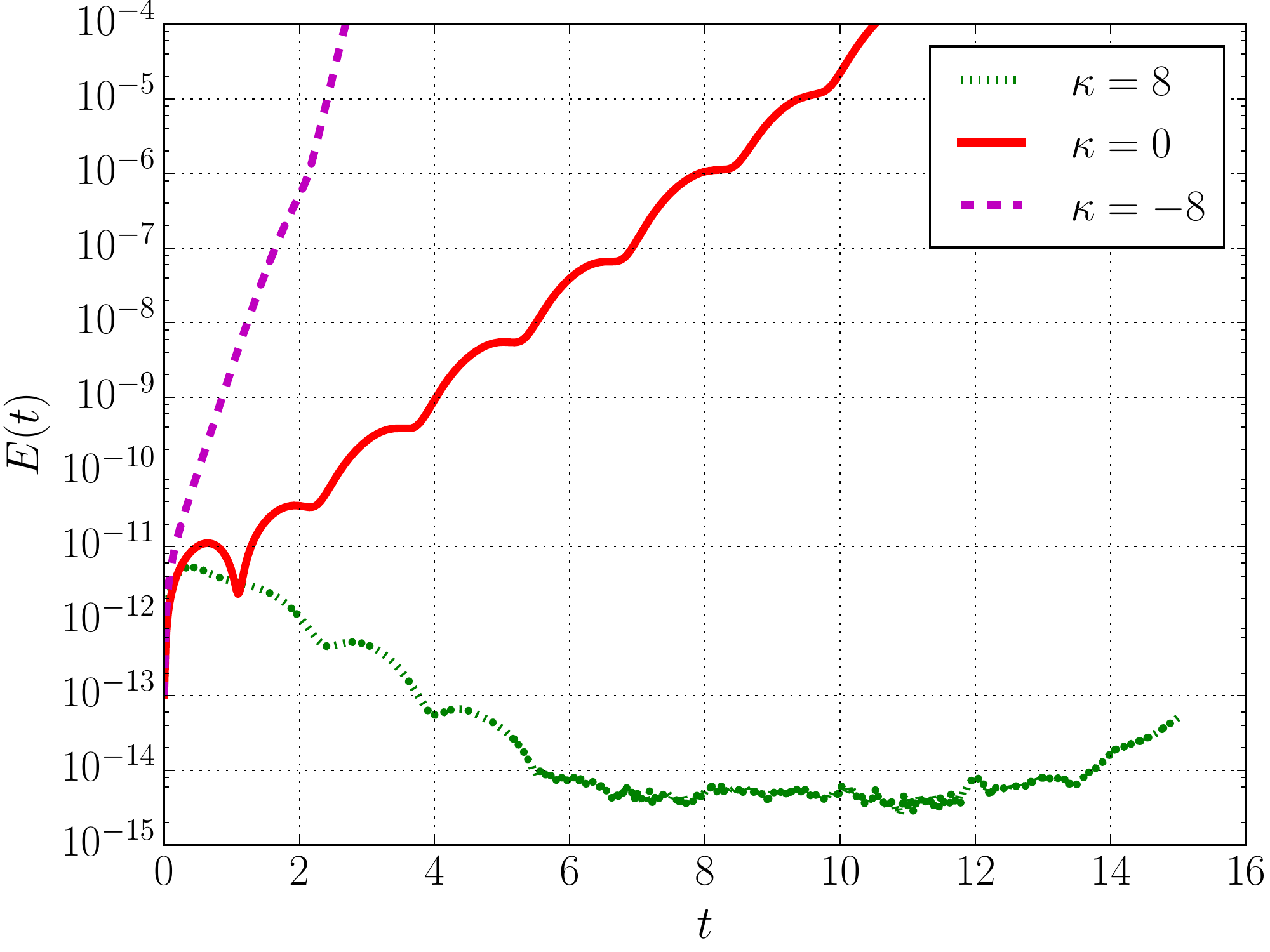}  
    \caption{The constraint violation error $E(t)$ for fixed Runge-Kutta time step $dt = 0.02$ and different values of $\kappa$.} \label{fig:error_comparison_k}        
    \end{minipage} 
    \hfill  
    \begin{minipage}{0.49\linewidth}
    \centering      
    \includegraphics[width=\textwidth]{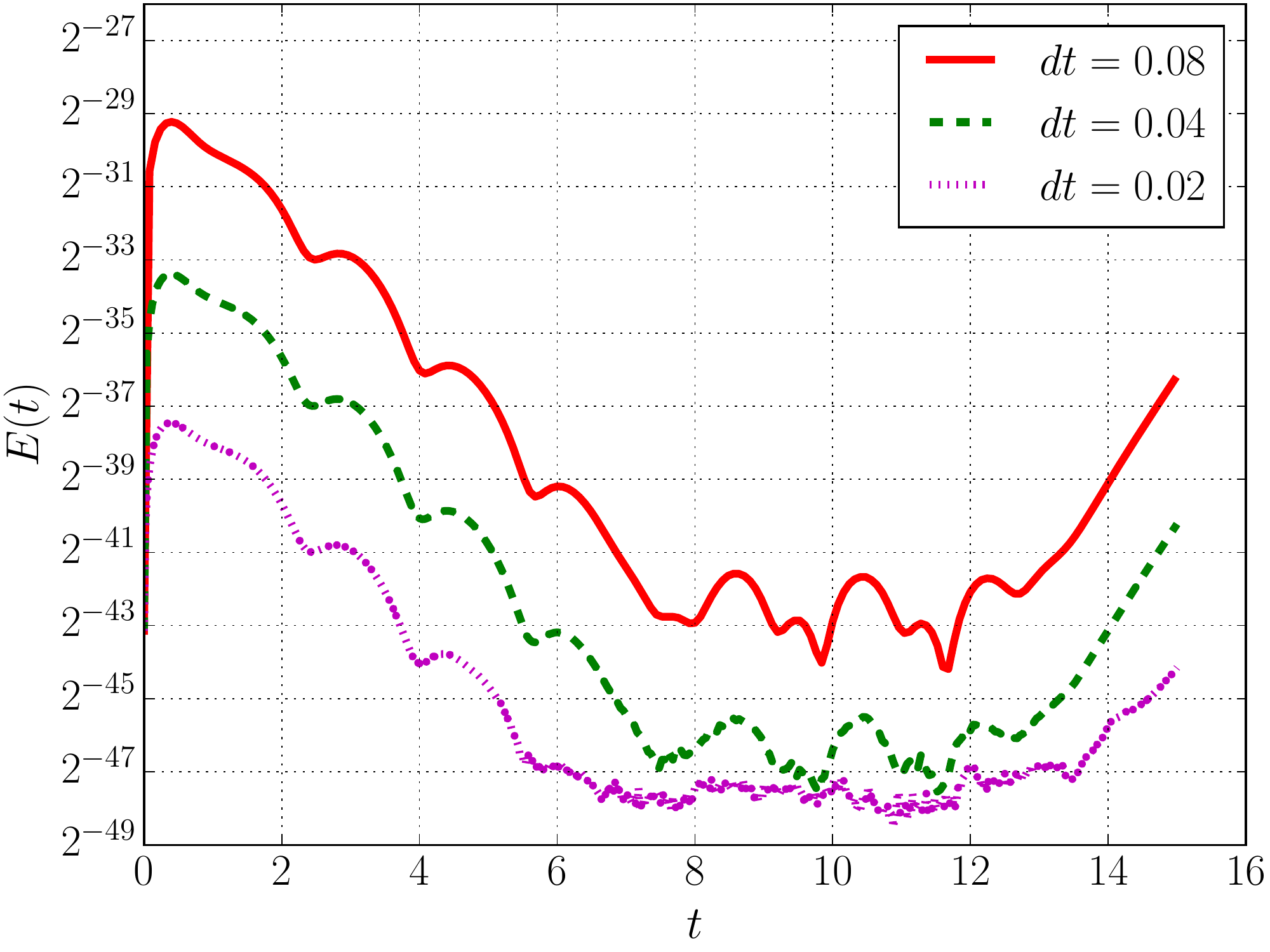}    
    \caption{The constraint violation error $E(t)$ for various Runge-Kutta time steps $dt$ and $\kappa=8$.} \label{fig:error_comparison_tol}  
  \end{minipage}       
\end{figure}

In order to check numerically the validity of the above statements let us consider the \textit{constraint error} as\footnote{We have excluded $\mathcal{D}_2(t)$ from this definition because $\mathcal{D}_1(t,\theta)=\overline{\mathcal{D}_2(t,\theta)}=\mathcal{D}_2(t,\theta)$ in Gowdy symmetry.} 
\begin{equation*}
E(t):= \sqrt{ \| \: \mathcal{D}_0(t,\theta)  \:  \|_{L^2(\St)}^{2} + \| \: \mathcal{D}_1(t,\theta)  \:  \|_{L^2(\St)}^{2}  } \: ,
\end{equation*}
where the norm $\| \: .  \:  \|_{L^2(\St)}$ is numerically computed by using \Eqref{ec:L2_norm}. In \Figsref{fig:error_comparison_k} and \ref{fig:error_comparison_tol}, we plot this error obtained for different values of  $\kappa$. Here, we have used the initial data family which we will describe in \Sectionref{sec:familyIDN} and pick  $\epsilon = -10^{-4}$, $C=0$ and $\ell=2$. In order to calculate $E(t)$ in these figures, we calculate the numerical solution of the evolution equations with constraint damping terms corresponding to these data using the pseudo-spectral method described in \cite{Beyer:2016fc} with the Runge-Kutta method as time integrator and the axial symmetric spin weighted transform for computing the spatial derivatives (more details on the numerical infrastructure are given below). 
As  expected, the error for $\kappa=-8$ grows exponentially whereas for $\kappa=8$ it is bounded. Thus, from now on, we will keep this value for  all the following numerical calculations in this paper.

\paragraph{Numerical infrastructure} 

The main difficulty for the numerical treatment of tensorial equations on manifolds with spherical topology is the fact that these cannot be globally covered  by a single regular coordinate patch. In the literature this problem is commonly known as the \textit{pole problem} because in standard polar coordinates for \St these issues appear at the poles. Based on the previous works \cite{Beyer:2014bu,Beyer:2015bv}, we introduced in \cite{Beyer:2016fc}  a pseudo-spectral infrastructure to overcome this issue numerically. It consists in using the  {spin-weight formalism} for  expressing  tensor components in terms of {spin-weighted spherical harmonics}, which are a generalization of the well known spherical harmonics \cite{Penrose:1984tf}; see \Appendixref{Sec:swsh}. This allows us to work with polar coordinate representations of \Eqsref{eq:geroch_einstein_equations}  that do not suffer from any polar singularity. This becomes manifest when all the spatial derivatives are replaced by the eth-operators in \Eqsref{eq:def_eths}. 

As mentioned earlier, we shall exclusively restrict to Gowdy symmetric models which implies axial symmetry for all the fields in \Eqsref{eq:geroch_einstein_equations}. 
For the numerical treatment of such fields, we have introduced the one-dimensional variant of the spin-weighted transform introduced by Huffenberger  and Wandelt \cite{Huffenberger:2010hh}, which we  call \textit{axially symmetric spin-weighted transform} in \cite{Beyer:2016fc}. Our numerical infrastructure is therefore a pseudo-spectral scheme based on the method of lines where the temporal integration is carried out by certain Runge-Kutta integrators.

%%%%%%%%%%%%%%%%%%%%%%%%%%%%%%%%%%%%%%%%%%%%%%%%%%%%%%%%%%%%%%%%%%%%%%%%%%%%%%%%%%%%%%%%%%%%%%%%%%%%%%%%%%%%%%%%%%%%%%%%%%%%%%%%%
%%%%%%%%%%%%%%%%%%%%%%%%%%%%%%%%%%%%%%%%%%%%%%%%%%%%%%%%%%%%%%%%%%%%%%%%%%%%%%%%%%%%%%%%%%%%%%%%%%%%%%%%%%%%%%%%%%%%%%%%%%%%%%%%%
\subsection{Constraints and initial data}
\label{sec:familyIDN}

\paragraph{Formulation of the constraints and choice of free data}
As explained in \Sectionref{sec:formulationeqs}, initial data for the evolution equations of Einstein's equations must satisfy, first, the gauge constraint
\[0=h^{\rho\sigma}(\bar\Gamma^{\nu}{}_{\rho\sigma}-\Gamma^{\nu}{}_{\rho\sigma})+f^\nu,
\]
recall \Eqref{eq:constraint1}.
Second, we must respect the Hamiltonian and Momentum constraints associated with \Eqref{eq:geroch_einstein_equations} which take the form 
\begin{equation}\label{eq:ADMconstrints}
\begin{split}
^{(2)}R + \mathcal{K}^{2}  - \mathcal{K}_{ ik } \mathcal{K}^{ik} -  \dfrac{2\Lambda}{\psi}   &=  2 \rho ,\\
D_k ( \mathcal{K}^{ki} - \gamma^{ki} \mathcal{K} ) &= j^{i}.
\end{split}
\end{equation}
In this subsection, we use abstract indices $i, j, k, \ldots$ to represent two-dimensional purely spatial fields. 
Notice that in this subsection only, we use all the symbols $\delta$, $\phi$ etc., which we had introduced for fields on $S$ before, now to denote the restriction of these quantities to any $t=\mathrm{const}$-surface. The values of their time derivatives are denoted as $\dot\delta$, $\dot\phi$ etc. The quantity $\gamma_{ik}$ above is the induced $2$-metric (see \Eqref{eq:generalinducedmetric}) and  $D_k$ the corresponding covariant derivative. $^{(2)}R$ is the scalar curvature associated with $\gamma_{ik}$ and $\mathcal{K}_{ik}$ represents the extrinsic curvature with $\mathcal{K}=\mathcal{K}^i {}_{i}$. 
If $T_{ab}$ is the energy-momentum tensor of the matter source in the $2+1$-Einstein equations in \Eqsref{eq:geroch_einstein_equations}, then
\[\rho = n_{a} n_{b} T^{ab}, \quad j^{i} = -\gamma^{i}{}_{a} n_{b} T^{ab},\]
and hence
\begin{eqnarray}
\label{eq:density}
{\rho} &=& \dfrac{ \dot \psi^2+ \dot \omega^2 +2|m^i\nabla_i\psi|^2  +2|m^i\nabla_i \omega|^2}{4 \delta \psi^2},\\
\label{eq:current}
j^1= j^2&=& -\dfrac{\dot\psi m^i\nabla_i\psi +\dot\omega m^i\nabla_i\omega}{2\sqrt{\delta}\, \psi^2},
\end{eqnarray}
Here the vector $j^i$ has been expressed in terms of the spatial frame $(\partial_1^i,\partial_2^i)=(m^i,\mbar^i)$; recall \Eqsref{eq:referenceframe} and \eqref
{eq:referenceframe3}. The corresponding spatial dual frame $(\omega^1_i, \omega^2_i )$ is defined as in \Eqref{eq:referenceframe2}.
Notice that \Eqsref{eq:generalinducedmetric} and \eqref{eq:conformaltimegauge} yield
\begin{equation}
\label{eq:gammaij}
\gamma_{ik} =  2 \delta \omega_{(i}^1 \omega_{k)}^2  + \phi \left(\omega_i^1 \omega_k^1 + \omega_i^2 \omega_k^2\right).
\end{equation}

Because the shift $\beta$ vanishes as a consequence of \Eqref{eq:conformaltimegauge}, the extrinsic curvature $\mathcal K_{ik}$ is proportional to the time derivative of $\gamma_{ik}$ and is therefore determined by $\dot\delta$ and $\dot\phi$.
The first step of finding a complete set of initial  data on the $t=0$-surface is to find the quantities
$\delta, \dot\delta, \phi, \dot\phi, \psi, \dot\psi, \omega, \dot\omega$
as solutions of the Hamiltonian and Momentum constraints. Once this is done we find initial data for
$f_0, f_1$
as a solution of the gauge constraint in a second step. Recall that all these quantities are assumed to be real and only depend on $\theta$.

We shall construct solutions of the Hamiltonian and Momentum constraints using the York-Lichnerowicz conformal decomposition; see \cite{Bartnik:2004wn} and references therein. We shall not describe the general procedure here (which, in two dimensions, is slightly different from the standard $3$-dimensional case), but restrict to the simple time symmetric case
\[\mathcal K_{ik}=0\]
in all of what follows. According to \Eqref{eq:gammaij} and the choice of vanishing shift, this is the case if and only if
\begin{equation}
  \label{eq:IDchoice3}
  \dot\delta=0,\quad\dot\phi=0.
\end{equation}
The momentum constraint is therefore satisfied if $j^{1}=j^{2}=0$. According to \Eqref{eq:current}, this is in particular the case if we pick
\begin{equation}
  \label{eq:choiceid1}
  \omega = 1 - \psi,\quad \dot\psi = \dot\omega.
\end{equation}

The Momentum constraint is now satisfied and hence we attempt to solve the Hamiltonian constraint next. Because the topology of the spatial slices is $\St$, we can assume that $\gamma_{ik}$ is initially conformal to the standard round unit two-sphere metric. 
According to \Eqref{eq:gammaij}, we can therefore pick
\[\phi=0,\]
which yields
\begin{equation*}\label{ec:comformal_trans}
\gamma_{ik} := \delta \mathring{ \gamma }_{ik} ,
\end{equation*}
where $\mathring{ \gamma }_{ik}$ represents the metric for the unit round two-sphere 
\begin{equation*}
\mathring{ \gamma }_{ik} =  2 \omega_{(i}^1 \omega_{k)}^2.
\end{equation*}
The quantity $\delta$ can therefore be considered as the conformal factor in the standard conformal decomposition of the Hamiltonian constraint.
We  express the two-dimensional Ricci scalar in the Hamiltonian constraint as 
\begin{equation}\label{ec:ricciforhamiltonian}
^{(2)}R = \delta^{-1} \left( \mathring{R} -  \mathring{D}_{i} \mathring{D}^{i} \ln \delta  \right) ,
\end{equation}
where $\mathring{R}=2$ is Ricci scalar of the unit round two-sphere. Replacing   frame derivatives by eth-operators by means of \Eqref{eq:ethm}, a straightforward calculation recasts the  Hamiltonian to the form
\begin{equation}\label{eq:hamiltonian}
\Delta_{\St} \delta=\bar{\eth} \eth\; \delta = 2 \delta -2\delta^2 \left(   \dfrac{\Lambda}{\psi} + \rho \right)  + \dfrac{ |\eth \delta|^2 }{\delta}.  
\end{equation}
See \Eqref{eq:DefDelta} for our definition of the Laplace operator $\Delta_{\St}$ on the $2$-sphere.
Using \Eqsref{eq:density} and \eqref{eq:choiceid1}, we get
\begin{equation*}\label{semihamiltonian}
\Delta_{\St} \delta = 2 \delta -\frac{2 \delta ^2  }{\psi }-\frac{\delta \dot\psi ^2}{\psi ^2}+\frac{|\eth \delta |^2 }{\delta } - \frac{\delta |\eth \psi|^2 }{\psi ^2}. 
\end{equation*}
If we now pick 
\begin{equation}
  \label{eq:IDchoice2}
  \psi = \delta^2,
\end{equation}
the only remaining free function is $\dot\psi$ in terms of which the Hamiltonian constraint becomes
\begin{equation}\label{ec:hamiltonian_numeical}
\Delta_{\St} \delta = -2+ 2 \delta -  \dfrac{\dot\psi^2}{ \delta^3} -\dfrac{3 | \eth \delta |^2 }{\delta }. 
\end{equation}
Note that for $\dot\psi=0$ and $\delta=\text{const}$, the trivial (but not the only) solution of \Eqref{ec:hamiltonian_numeical} is $\delta = 1$ which yields the initial data of the Nariai metric. Hence, in order to obtain \emph{perturbations} of the Nariai spacetime, we only have to provide non-zero functions $\dot\psi$ and solve numerically \Eqref{ec:hamiltonian_numeical}.
In the whole paper, we choose
\begin{equation}\label{num_initial_data1}
\dot\psi = \epsilon Y_{\ell} (\theta) + C Y_0(\theta), 
\end{equation}
where $\epsilon$ and $C$ are free real parameters  and $\ell$ is any fixed positive integer. We list the necessary information about the functions $Y_\ell$ in \Sectionref{Sec:swsh}. Observe that
$Y_0(\theta)={1}/(2\sqrt{\pi})$.

It only remains to provide initial data for $f_0$ and $f_1$ as solutions of the gauge constraint. 
Given the choices above, it turns out that
the gauge constraint is satisfied if and only if
\[f_0=0,\quad f_1 = f_2= -\eth{\delta}/(2 \sqrt{2}\delta).\]

Once all this is done, so, in particular, once we have solved \Eqref{ec:hamiltonian_numeical} with \Eqref{num_initial_data1}, our initial data set is complete and satisfies all the required constraints: the Hamiltonian constraint, the Momentum constraint, and, the gauge constraint.

\paragraph{Numerical method to solve the Hamiltonian constraint}
Now, we describe the basic idea for using a spectral implementation based on the spin-weighted spherical harmonics in the axi-symmetric case (no $\varphi$-dependence) for solving \Eqref{ec:hamiltonian_numeical} with \Eqref{num_initial_data1}. We follow the approach in \cite{Frauendiener:1999vr} for solving  non-linear elliptic equations. For more information about these kind of methods, the interested reader is referred to \cite{Fornberg:1998gv,orszag1980spectral} and references therein. 

Let us start by writing the right-hand side of \Eqref{ec:hamiltonian_numeical} as a non-linear function $f(\delta, \eth \delta )$ with spin-weight $0$. 
The idea is then to construct a sequence of linearized problems whose solutions hopefully converge to the solution of the non-linear problem. 
For the Richardson's iteration procedure this sequence of solutions $(\delta_n)$ is constructed 
by solving 
\begin{equation}\label{lienarized_hamiltonian_constraint}
\Delta_{\St} \; \zeta  -  \left( \dfrac{\partial f}{\partial \delta} \right)_n \zeta - \left( \dfrac{\partial f}{\partial \: \eth\delta} \right)_n \eth \; \zeta = - \left(  \Delta_{\St} \delta_n - f_n \right)
\end{equation}
for each $n=0,1,2,\ldots$ for some \keyword{initial guess} $\delta_0$ and then to set
\[\delta_{n+1} = \delta_{n} + \zeta.\]
We call $\zeta$  the \textit{correction factor}. 
The right-hand side of this equation is known as the \keyword{residual} $r_n$ at the step  $n$, that measures how well $\delta_n$ satisfies the equation at the step $n$. 

In our pseudo-spectral approach, we introduce suitable collocation points $\theta_1 , \ldots, \theta_N$, and impose  \Eqref{lienarized_hamiltonian_constraint} at those.
Using the properties of the eth-operators listed in \Eqsref{eq:eths}, this yields an algebraic linear system of equations for $N$ spectral coefficients of $\zeta$ when written in the spin-weighted spherical harmonics basis. 
We shall not discuss the details of the solvability of this linear system here. However, it is guaranteed to have a unique solution in each step if 
the coefficients $(\partial f/\partial\delta)_n$ and $(\partial f/\partial \: \eth\delta)_n$ satisfy certain algebraic conditions in each step. If this is the case, then the iteration converges quickly and  thereby allows us to construct accurate approximations of solutions of the nonlinear equation. 

\begin{figure}[t]
    \centering
     \includegraphics[width=0.49\textwidth]{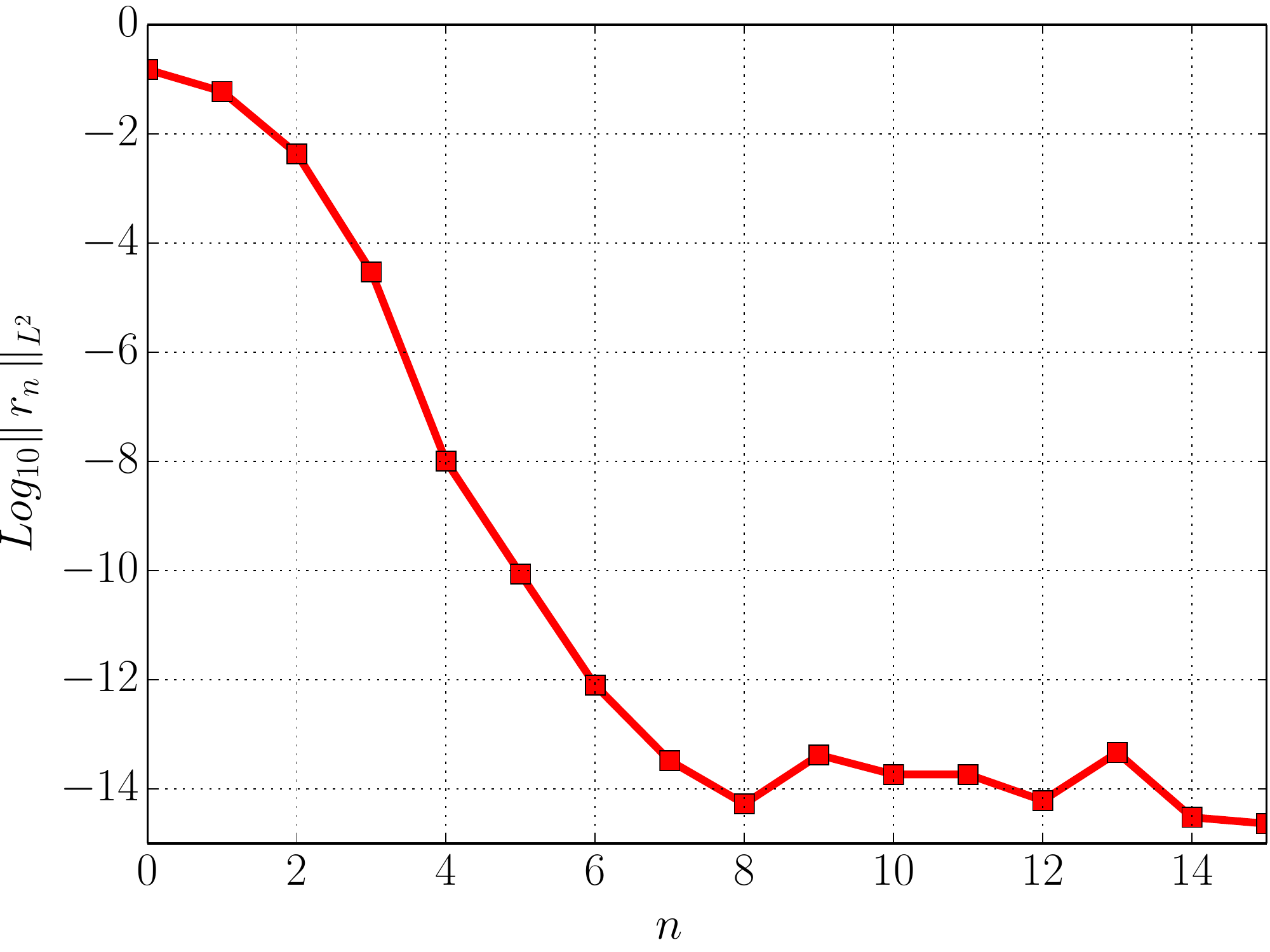}  
     \caption{Convergence of the numerical scheme for solving the Hamiltonian constraint.} \label{convergence_initial_data}    
 \end{figure}

In \Figref{convergence_initial_data}, choosing  $\epsilon=-10^{-4}$, $C=0$ and $\ell=2$,  we show the  behavior of the norm $ \| r_n \|_{L^2(\St)}$ as a function of $n$ which is numerically  approximated  by
\begin{equation}\label{ec:L2_norm}
\| \:  r_n  \:  \|_{L^2(\St)}^2 \approx  \dfrac{ 2 \pi^2 }{ N } \:  \sum \limits^{N}_{i=0} \: r_n^2.
\end{equation}
In this figure we observe  that the norm of $r_n$ decays rapidly until it reaches a satisfactory order of $\sim 10^{-14}$. 

\paragraph{Approximate analytic solutions of the Hamiltonian constraint}
The heuristic analytical approach in \Sectionref{sec:heuristicmode} below relies on the following \emph{analytic} approximations of solutions which is meaningful at least when the parameters $\epsilon$ and $C$ are small.

To this end, we shall now assume that the family of solutions $\delta$ of \Eqref{ec:hamiltonian_numeical} with \Eqref{num_initial_data1}  depends smoothly on the parameters $\epsilon$ and $C$ in a neighborhood of $\epsilon=C=0$. Then we express $\delta$ approximately as
\begin{equation}
  \label{eq:deltaansatz}
  \delta(\theta)=1
  +\epsilon\delta_{(1)}(\theta)+C\delta_{(2)}(\theta)
  +\epsilon^2\delta_{(3)}(\theta)
  +\epsilon C\delta_{(4)}(\theta)
  +C^2\delta_{(5)}(\theta)+\ldots,
\end{equation}
for some so far unknown functions $\delta_{(1)},\ldots,\delta_{(5)}$ which are assumed to be independent of $C$ and $\epsilon$. 
With this ansatz, we find that  \Eqsref{ec:hamiltonian_numeical} and \eqref{num_initial_data1} are satisfied up to cubic order in the parameters, if 
\begin{equation*}
  2\delta_{(1)}-\Delta_{\St}\delta_{(1)}=0,\quad
  2\delta_{(2)}-\Delta_{\St}\delta_{(2)}=0,
\end{equation*}
which is implied by the linear orders in $\epsilon$ and $C$ and which yields that
\begin{equation}
  \label{eq:blub1}
  \delta_{(1)}=\delta_{(2)}=0,
\end{equation}
and, if
\begin{equation}
  \label{eq:blubbbb}
  2\delta_{(3)}-\Delta_{\St}\delta_{(3)}=(Y_\ell)^2,\quad
  2\delta_{(4)}-\Delta_{\St}\delta_{(4)}=\frac{Y_\ell}{\sqrt{\pi}},\quad
  2\delta_{(5)}-\Delta_{\St}\delta_{(5)}=\frac {Y_0}{2\sqrt{\pi}},
\end{equation}
where \Eqref{eq:blub1} and $Y_0={1}/(2\sqrt{\pi})$ have been used to simplify these equations.
It is well known that for any PDE of the form
  \[p u(\theta)-\Delta_{\St}u(\theta)=f(\theta)=\sum_{k=0}^\infty f_k Y_k(\theta)\]
  defined on $\St$
given by any smooth source term function $f$ and any non-negative integer $p$, the uniquely determined solution is
\[u(\theta)=\sum_{k=0}^\infty\frac{f_k}{p+k (k+1)} Y_k(\theta).\]
Regarding \Eqsref{eq:blubbbb}, this implies that
\begin{align*}
  \delta_{(3)}=\sum_{k=0}^{2\ell}\frac{a_{\ell,k}}{2+k (k+1)} Y_k,\quad
  \delta_{(4)}=\frac{1}{(2+\ell (\ell+1))\sqrt\pi} Y_\ell,\quad
  \delta_{(5)}=\frac {Y_0}{4\sqrt{\pi}}.
\end{align*}
The coefficients $a_{\ell,k}$ here are defined implicitly by the equation
\begin{equation}
  \label{eq:ClebschGordonQuadr}
  (Y_{\ell})^2=\sum_{k=0}^{2\ell} a_{\ell,k} Y_k
\end{equation}
which can therefore be calculated explicitly from the well-known Clebsch-Gordon coefficients \cite{Sakurai:1994modern}.
When we combine all this with \Eqref{eq:deltaansatz} we find
\begin{equation}
  \label{eq:approxsolN}
    \delta=1
    %+\epsilon\delta_{(1)}(\theta)+C\delta_{(2)}(\theta)
    +\sum_{k=0}^{2\ell}\frac{a_{\ell,k}\epsilon^2 }{2+k (k+1)} Y_k
    +\frac{\epsilon C }{\sqrt\pi (2+\ell (\ell+1))} Y_\ell
    +\frac {C^2}{4\sqrt\pi} Y_0+\ldots.
  \end{equation}
This is an approximation of solutions of \Eqref{ec:hamiltonian_numeical} with \Eqref{num_initial_data1} which is expected to be valid for small values of the parameters $C$ and $\epsilon$.

\begin{figure}[t]  
  \centering
  \includegraphics[width=0.49\textwidth]{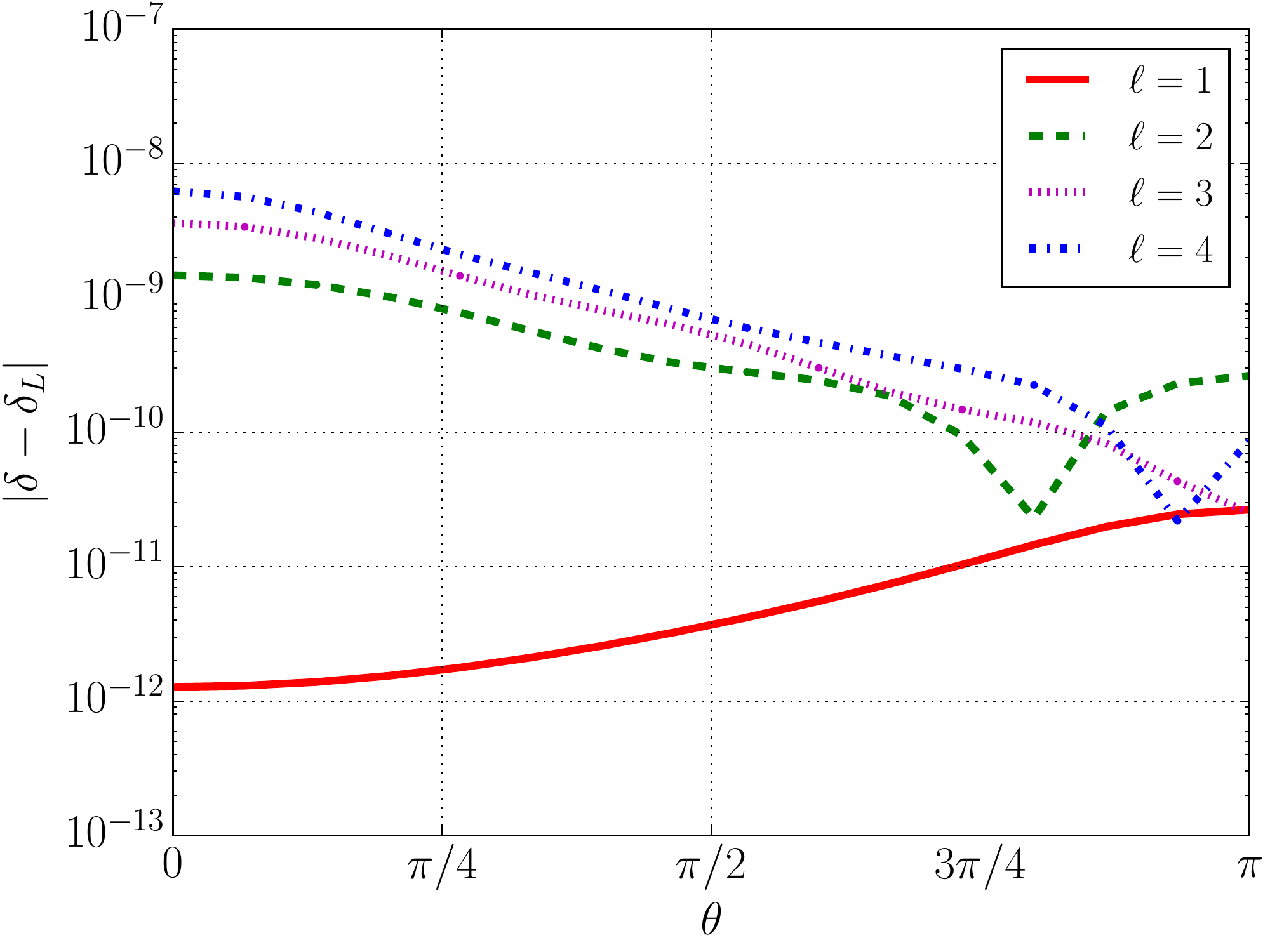}
  \hfill
  \includegraphics[width=0.49\textwidth]{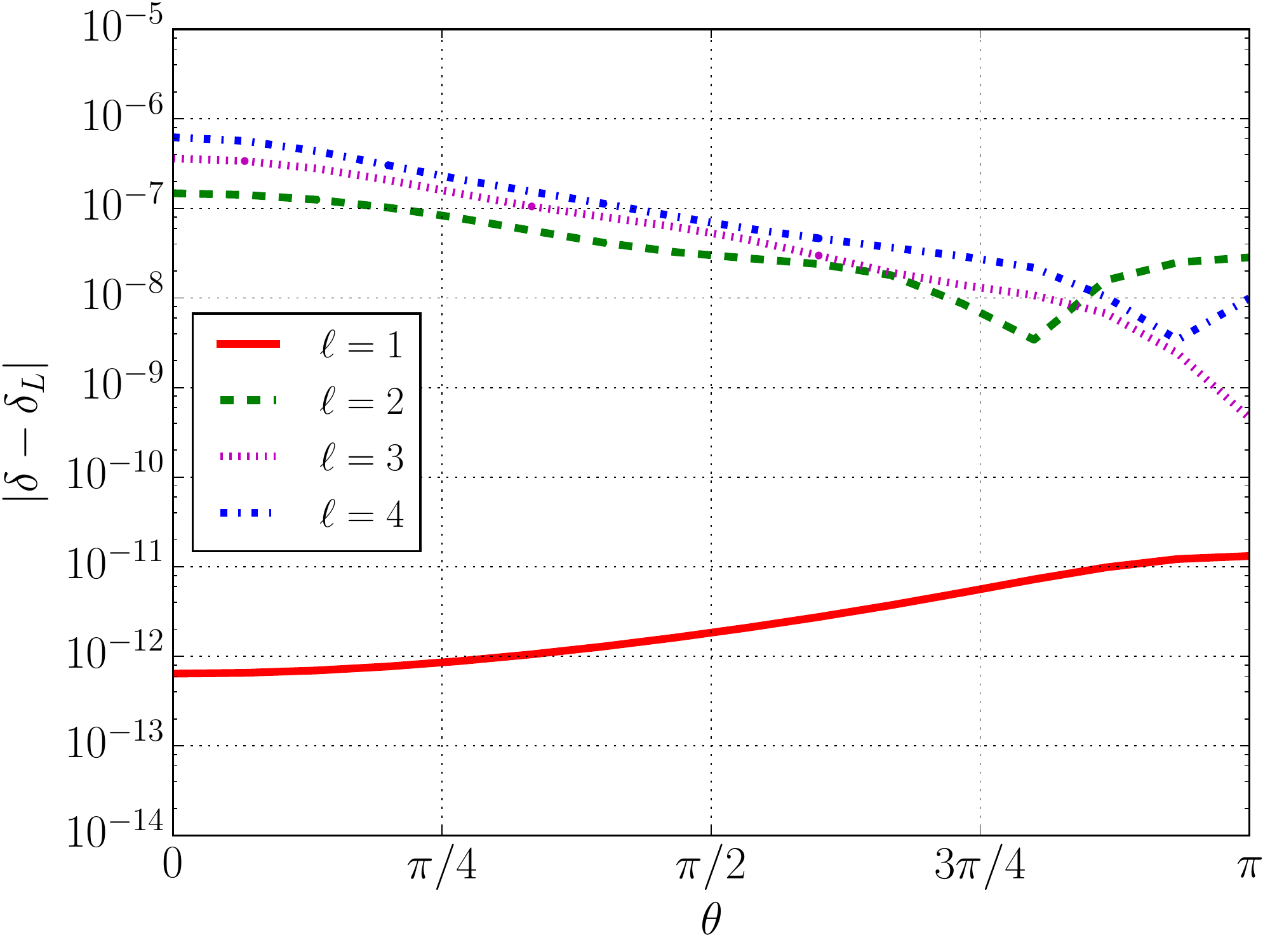}
  \caption{Difference of the numerical solution $\delta$ of the Hamiltonian constraint and the analytic expression $\delta_L$ given by \eqref{eq:approxsolN} for $\epsilon=10^{-4}$ and $\epsilon=10^{-3}$, $C=0$ and various values of $\ell$.}
  \label{fig:IDError}
\end{figure}

In \Figref{fig:IDError} we provide  numerical evidence which supports the claim that \Eqref{eq:approxsolN} is a good approximation of solutions of the Hamiltonian constraint in many of the cases of interest.

%%%%%%%%%%%%%%%%%%%%%%%%%%%%%%%%%%%%%%%%%%%%%%%%%%%%%%%%%%%%%%%%%%%%%%%%%%%%%%%%

\section{Analysis and results}
\label{sec:criticalNariai}

\subsection{Heuristic mode analysis}
\label{sec:heuristicmode}

Our interpretation of our numerical results and conclusions below are based on a \keyword{heuristic mode analysis technique} which we shall discuss first now. 
Recall that the unknowns of our dynamical equations are $\delta$, $\phi$, $\psi$, $\omega$, $f_0$ and $f_1$ which are all real quantities of spin-weight $0$, $2$, $0$, $0$, $0$ and $1$, respectively, depending only on $t$ and $\theta$. In order to facilitate the following analysis we define
\begin{equation}
  \label{eq:defdeltaS}
  \delta_*:=\psi^{-1}\delta, \quad \phi_*:=\psi^{-1}\phi,
\end{equation}
which are related to the \textit{physical} $2+1$-metric $\hat h_{ab}$ as
\begin{equation}
  \label{eq:physmetric}
  \hat h_{ab}=-\delta_*\omega_a^0 \omega_b^0 +2 \delta_* \omega_{(a}^1 \omega_{b)}^2  + \phi_* (\omega_a^1 \omega_b^1 + \omega_a^2 \omega_b^2)
\end{equation}
in the gauge \Eqref{eq:conformaltimegauge}; cf.\ \Eqsref{ec:21metric} and \eqref{eq:conformal_metric}.
Moreover, we define
\begin{equation}
  \label{eq:defpsiS}
  \psi_*:= \text{sech}^2 t \ \psi , \quad\omega_*:= \psi^{-1} \omega  ,
\end{equation}
and set
  \begin{equation*}
    u_* := \left(  \delta_* , \phi_*, \psi_*,  \omega_*, f_{0} ,  f_{1}  \right).
  \end{equation*}
All the components of $u_*$ -- and hence in particular the quantity $\delta_*$ which we shall mostly focus on in the following -- can be decomposed using spin-weighted spherical harmonics of appropriate spin-weight (see \Sectionref{Sec:swsh}), for example
  \begin{equation*}
    \delta_*(t,\theta)=\sum_{l=0}^\infty \delta_{*,l}(t) Y_l(\theta).
  \end{equation*}
Consistently with this, we write the collection of the $l$-th coefficients of all those components of $u_*$ for which these are defined schematically as $u_{*,l}$.
We shall refer to this  as the \textit{mode decomposition} of $u_*$ and $\delta_*$, respectively. The $l=0$-mode  $u_{*,0}$, so, in particular, $\delta_{*,0}$ will often be called \textit{fundamental mode}. We shall also often write 
\[u_{*h}=u_{*0},\quad \delta_{*h}=\delta_{*0},\]
in order to emphasize that these are the relevant modes in the spatially \textit{homogeneous case}. For the Nariai spacetime, we write $u_*=u_{*N}$ and $\delta_*=\delta_{*N}$ with
\begin{equation}
  \label{eq:deltaSNariai}
  \delta_{*N}=1\quad\Rightarrow\quad \delta_{*Nh}=2\sqrt{\pi} \quad\text{and}\quad\delta_{*N,l}=0,\quad\text{for all $l=1,2,\ldots$.}
\end{equation}

Using this, the evolution equation for $\delta_*$ can be written schematically as
\begin{equation}
  \label{eq:formallinevoleqpre}
  \ddot\delta_{*,l}(t)=G_l(t,u_{*,0}, u_{*,1}\ldots,\dot u_{*,0},\dot u_{*,1},\ldots).
\end{equation}
We may rewrite this 
for each $l=0,1,\ldots$ as
\begin{equation}
  \label{eq:formallinevoleq}
  \ddot\delta_{*,l}(t)+A_l (\dot\delta_{*,l}(t)-\dot\delta_{*N,l}(t)) +B_l (\delta_{*,l}(t)-\delta_{*,l}(t))=F_l(t,u_{*,0}, u_{*,1}\ldots,\dot u_{*,0},\dot u_{*,1},\ldots),
\end{equation}
where 
\begin{align*}
  A_l&:=-\left.\frac{\partial G_l}{\partial\dot\delta_{*,l}}\right|_{\substack{u_{*}=u_{*N},t=0}},
  \quad
  B_l:=-\left.\frac{\partial G_l}{\partial\delta_{*,l}}\right|_{\substack{u_{*}=u_{*N},t=0}},\\
  F_l&:=G_l+A_l  (\dot\delta_{*,l}-\dot\delta_{*N,l}) +B_l (\delta_{*,l}-\delta_{*N,l}).
\end{align*}
We emphasize that \Eqref{eq:formallinevoleq} just an algebraic manipulation of \Eqref{eq:formallinevoleqpre}. Also, it should be clear that similar decompositions can be performed for any of the other components of $u_*$. 
In any case, a lengthy  calculation now reveals that
\[A_l=0,\quad B_l=l(l+1)-2,\]
for all $l=0,1,\ldots$, and hence that
\begin{equation}
  \label{eq:formallin}
  \ddot\delta_{*,l}(t)+(l(l+1)-2) (\delta_{*,l}(t)-\delta_{*N,l}(t))=F_l(t,u_{*,0}, u_{*,1}\ldots,\dot u_{*,0},\dot u_{*,1},\ldots).
\end{equation}

Now, suppose we are in a regime where $|F_l|$ is negligible in comparison to the other terms in \Eqref{eq:formallin} and that the dynamics is therefore dominated by the left-hand side. 
Then, this equation together with \eqref{eq:deltaSNariai} implies
\begin{align}
  \label{eq:linsol1}
  \delta_{*,0}&\approx 2\sqrt{\pi}+(\left.\delta_{*,0}\right|_{t=0}-2\sqrt{\pi}) \cosh{\sqrt{2}\,t}+\left.\dot\delta_{*,0}\right|_{t=0} \sinh{\sqrt{2}\,t},\\
  \label{eq:linsol2}
  \delta_{*,1}&\approx \left.\delta_{*,1}\right|_{t=0} +\left.\dot\delta_{*,1}\right|_{t=0} t,\\
  \label{eq:linsol3}
  \delta_{*,l}&\approx \left.\delta_{*,l}\right|_{t=0} \cos{\sqrt{l(l+1)-2}\,t}+\left.\dot\delta_{*,l}\right|_{t=0} \sin{\sqrt{l(l+1)-2}\,t},
\end{align}
where $l\ge 2$. Hence, in this regime, the $l=0$ mode is in general unstable (in fact, this is the heuristic explanation for the before-mentioned instability of the Nariai solution in the class of homogeneous spacetimes) while the $l\ge 2$-modes are all oscillatory. The $l=1$-mode is ``somewhere in between''.

Before we continue, we wish to emphasize that the way the approximation \Eqsref{eq:linsol1} -- \eqref{eq:linsol3} is not a \textit{complete} linearization of the evolution equations around the Nariai spacetime. It is therefore questionable whether \Eqsref{eq:linsol1} -- \eqref{eq:linsol3} are useful in any sense. In any case, our numerical experiments below show that the rather simplistic description above turns out to be sufficient as a basic for our main results. 

Recall now that \Eqref{eq:approxsolN} is an approximation of the solution $\left.\delta\right|_{t=0}$ of the Hamiltonian constraint for our particular family of initial data. This approximation is expected to be valid for small parameter values $C$ and $\epsilon$.
Let us now use \Eqref{eq:approxsolN} to express \Eqsref{eq:linsol1} -- \eqref{eq:linsol3} in terms of the initial data parameters $\ell$, $\epsilon$ and $C$.
First, we see that \Eqsref{eq:defdeltaS}, \eqref{eq:IDchoice2} and \eqref{eq:IDchoice3} yield
\[\left.\delta_*\right|_{t=0}=\left.\frac 1{\delta}\right|_{t=0},\quad \left.\dot\delta_*\right|_{t=0}=\left.-\frac{\dot\psi}{\delta^3}\right|_{t=0}.\]
\Eqsref{eq:approxsolN} and \eqref{num_initial_data1} therefore give us the following result
\begin{align}
  \label{eq:initialdataexpr1}
  \left.\delta_*\right|_{t=0}&=1-\sum_{k=0}^{2\ell}\frac{a_{\ell,k}\epsilon^2 }{2+k (k+1)} Y_k
    -\frac{\epsilon C }{\sqrt\pi (2+\ell (\ell+1))} Y_\ell
    -\frac {C^2}{4\sqrt\pi} Y_0+\ldots,\\
  \label{eq:initialdataexpr2}
  \left.\dot\delta_*\right|_{t=0}&=-C Y_0-\epsilon Y_\ell+\ldots.
\end{align}
Now it turns out that in our applications, $C$ is typically much smaller than $\epsilon$. In fact, $C$ is often of the order $\epsilon^2$ (as justified below). When we combine \Eqsref{eq:linsol1} -- \eqref{eq:linsol3} with \Eqsref{eq:initialdataexpr1} and \eqref{eq:initialdataexpr2} and only keep terms of order $\epsilon$, $\epsilon^2$ and $C$, we get
\begin{align}
  \label{eq:linsol31N}
  \delta_{*,0}&\approx 2\sqrt{\pi}-\frac 12{a_{\ell,0}}\epsilon^2 \cosh{\sqrt{2}\,t}-C\sinh{\sqrt{2}\,t},\\
  \delta_{*,1}&\approx \frac 14 {a_{\ell,1}}\epsilon^2 -\epsilon d_{\ell,1} t,\\
  \label{eq:linsol32N}
  \delta_{*,l}&\approx -\frac{a_{\ell,l}\epsilon^2 }{2+l (l+1)} \cos{\sqrt{\ell(\ell+1)-2}\,t}-\epsilon d_{\ell,l}\sin{\sqrt{\ell(\ell+1)-2}\,t},
\end{align}
for all $l=2,\ldots,2\ell$ provided $\ell\ge 1$. Here we use  the notation
\[d_{i,k}:=
  \begin{cases}
  1 & i=k,\\
  0 & i\not= k.
  \end{cases}
\]

The approximate description \Eqsref{eq:linsol31N} -- \eqref{eq:linsol32N} of the dynamics of the quantity $\delta_*$ can of course be expected to hold only for small values of the initial data parameters $\epsilon$ and $C$ (i.e., close to the exact Nariai spacetime given by $\epsilon=C=0$) and only for short times $t$ close to the initial time $t=0$. As long as this approximation holds, it suggests that the criticality of the cosmological models, i.e., the borderline between collapse and expansion globally in space, is mainly governed by the fundamental mode $\delta_{*,0}$ because all other modes are bounded if $\ell\ge 2$. Moreover, for any choice of $\epsilon\in\R$, the \emph{critical value} of $C$, i.e., the value when $\delta_{*0}$ is exactly at the borderline in this approximation according to \Eqref{eq:linsol31N}, should be close to
\begin{equation}
  \label{eq:lincritC}
  C_{crit}=-\frac 12{a_{\ell,0}}\epsilon^2=-\frac 1{4\sqrt\pi}\epsilon^2,
\end{equation}
where we use that $a_{\ell,0}=1/(2\sqrt\pi)$ for all $\ell\ge 0$; recall the definition of $a_{\ell,l}$ by \Eqref{eq:ClebschGordonQuadr}. In our applications, we typically set $\epsilon=-10^{-\kappa}$ for some positive integer $\kappa$, which therefore yields
\begin{equation}
  \label{eq:lincritC2} 
  C_{crit}\approx -1.4\times 10^{-2\kappa-1}.
\end{equation}
Later we provide numerical evidence that the actually critical value of $C$ for solutions of the fully nonlinear equations is indeed somewhat close to \Eqref{eq:lincritC}.

If we now choose any $\ell\ge 2$ and pick initial data parameters $\epsilon$ and $C$ close to the actual critical values of the fully nonlinear problem for which all modes are expected to be bounded,  the oscillatory nature of the modes $\delta_{*,l}$ with $l=2,\ldots,2\ell$ suggested by \Eqref{eq:linsol32N} should dominate the dynamics. According to \Eqref{eq:linsol32N} the oscillation period is independent of $l$ (so long as $l$ is between $2$ and $2\ell$) and is given by
\begin{equation}
  \label{eq:linoscillationperiodN}
  T_L=\frac{2\pi}{\sqrt{\ell(\ell+1)-2}}.
\end{equation}
The phase and the amplitudes of the oscillations however depend significantly on $l$. If $l=\ell$ the amplitude is proportional to $\epsilon$ (in leading order), while it is proportional to $\epsilon^2$ for $l\not=\ell$; moreover there is a phase shift of approximately $\pi/2$ between these two kinds of modes. In the next subsections, we present numerical evidence which supports all  claims in this subsection.

\subsection{Numerical evidence for Result~1: Existence of critical models}

\begin{figure}[t]
  \centering
      \includegraphics[width=0.49\textwidth]{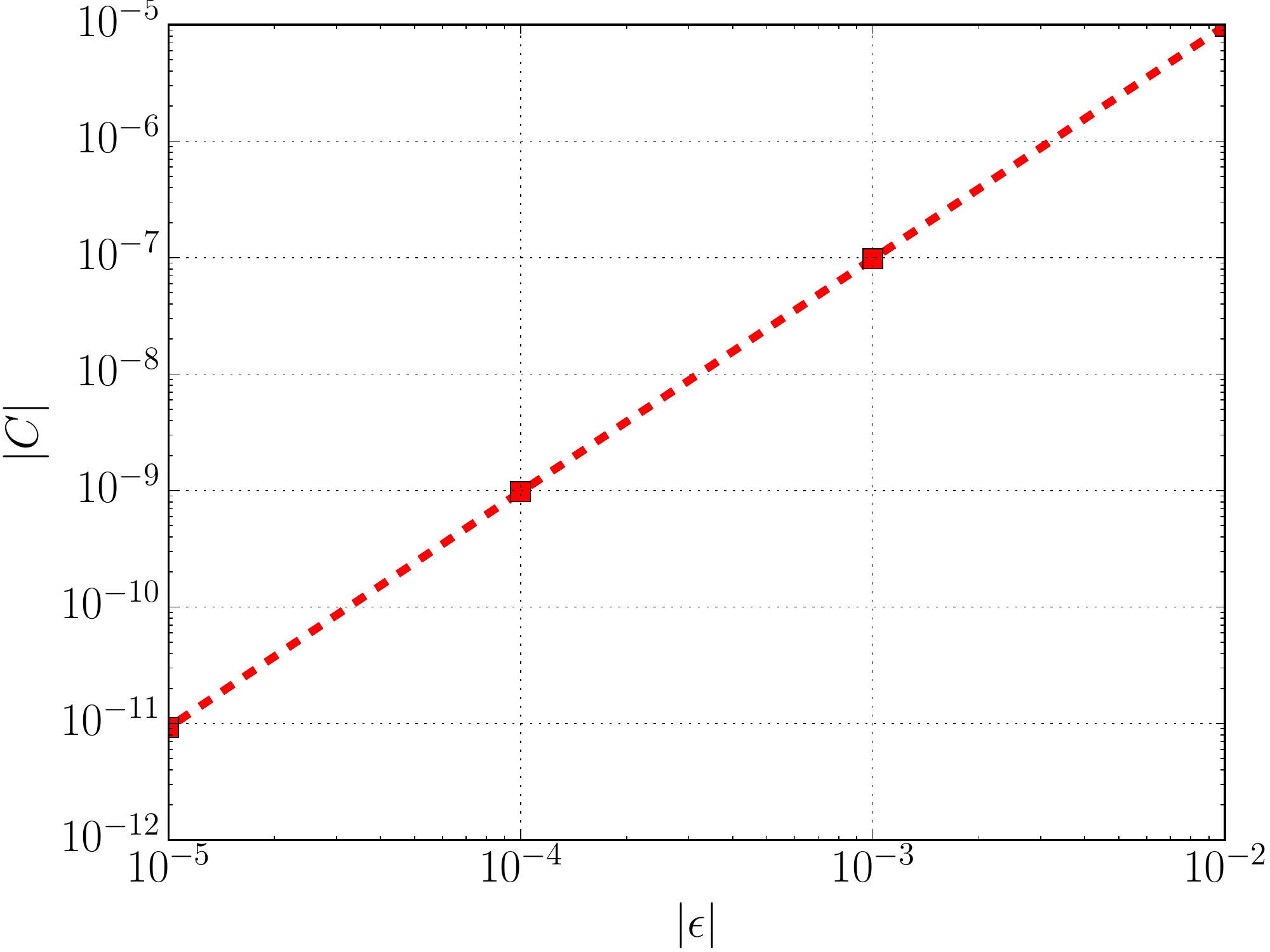}  
     \caption{The actual critical value $C_{crit}$ as a function of $\epsilon$ for $\ell=2$. The data points are taken from \Figref{fig:Different_epsilons}. The picture demonstrates that $C_{crit}\sim\epsilon^2$.} \label{fig:epsilons_and_Cs}
\end{figure}

\begin{figure}[t]
    \centering     
      \includegraphics[width=0.49\textwidth]{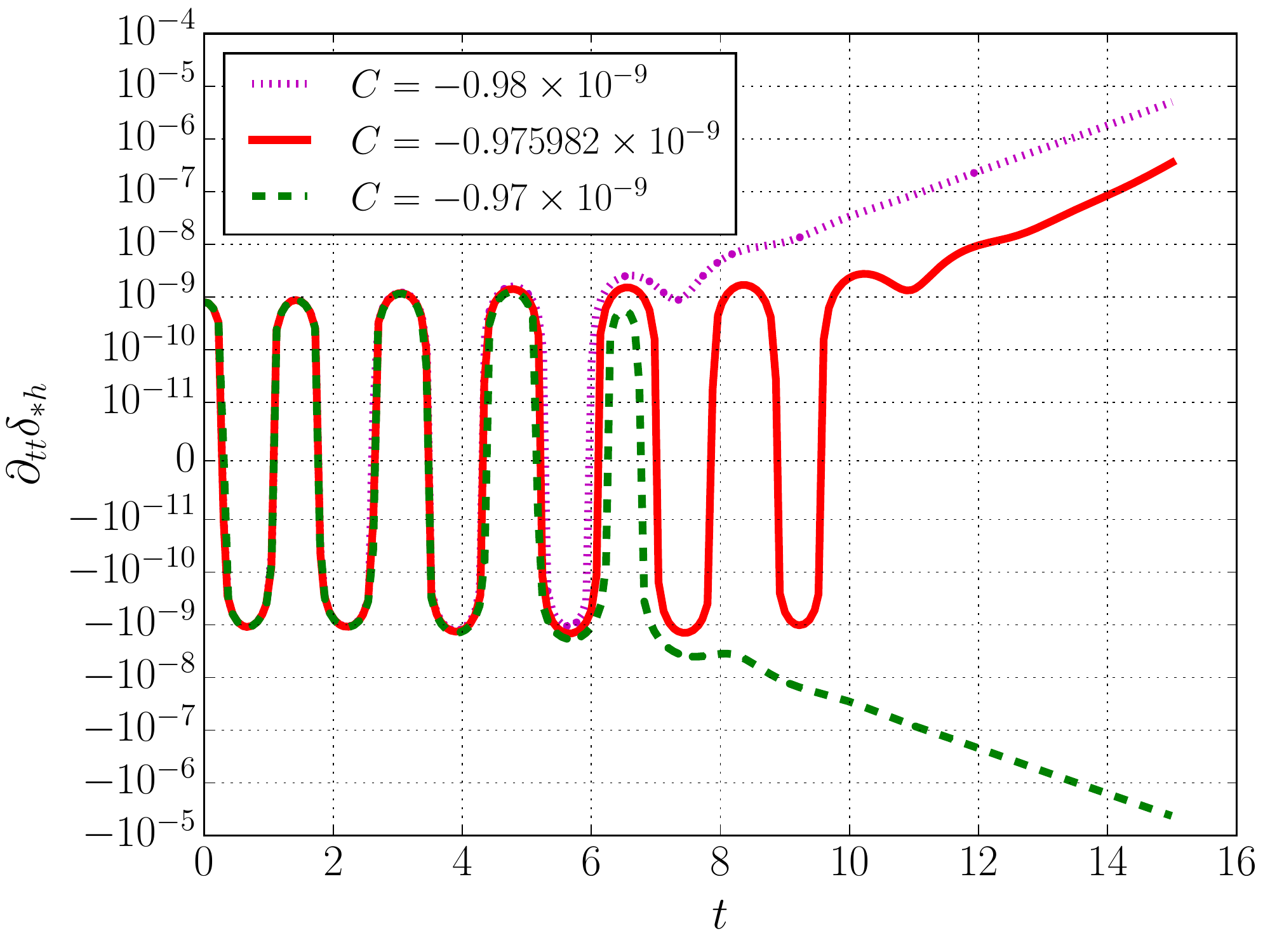}  
     \caption{Finding the critical solution with the algorithm in the text for $\ell=2$ and $\epsilon=-10^{-4}$.} \label{fig:CriticalBehaviour}
\end{figure}

\begin{figure}[t]
  \begin{minipage}{0.49\linewidth}
    \centering
     \includegraphics[width=\textwidth]{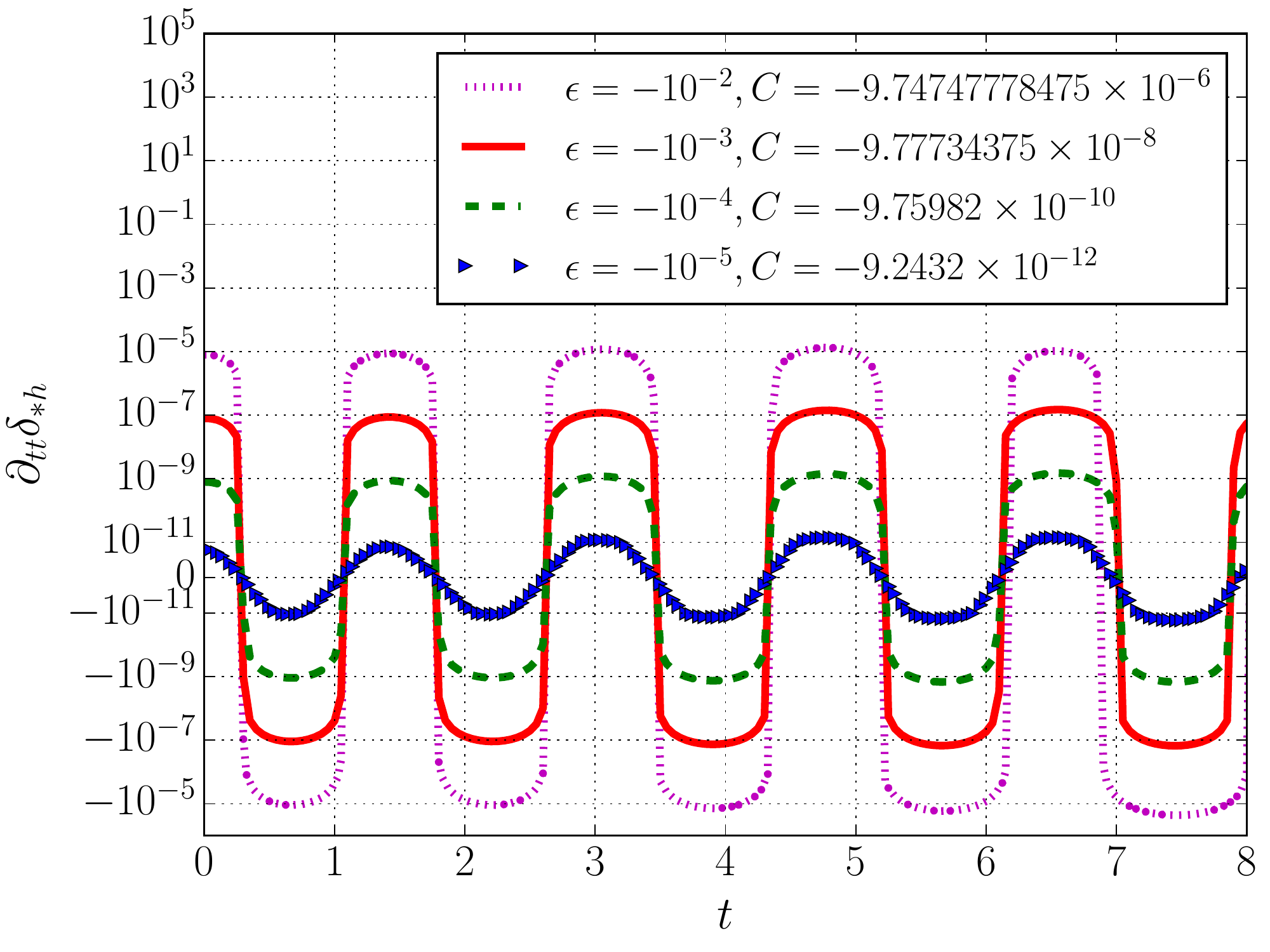}  
     \caption{The fundamental mode of our best numerical approximations of the critical solutions for various values of $\epsilon$ and fixed $\ell=2$.} \label{fig:Different_epsilons}    
  \end{minipage}  
  \hfill
  \begin{minipage}{0.49\linewidth}
    \centering
     \includegraphics[width=\textwidth]{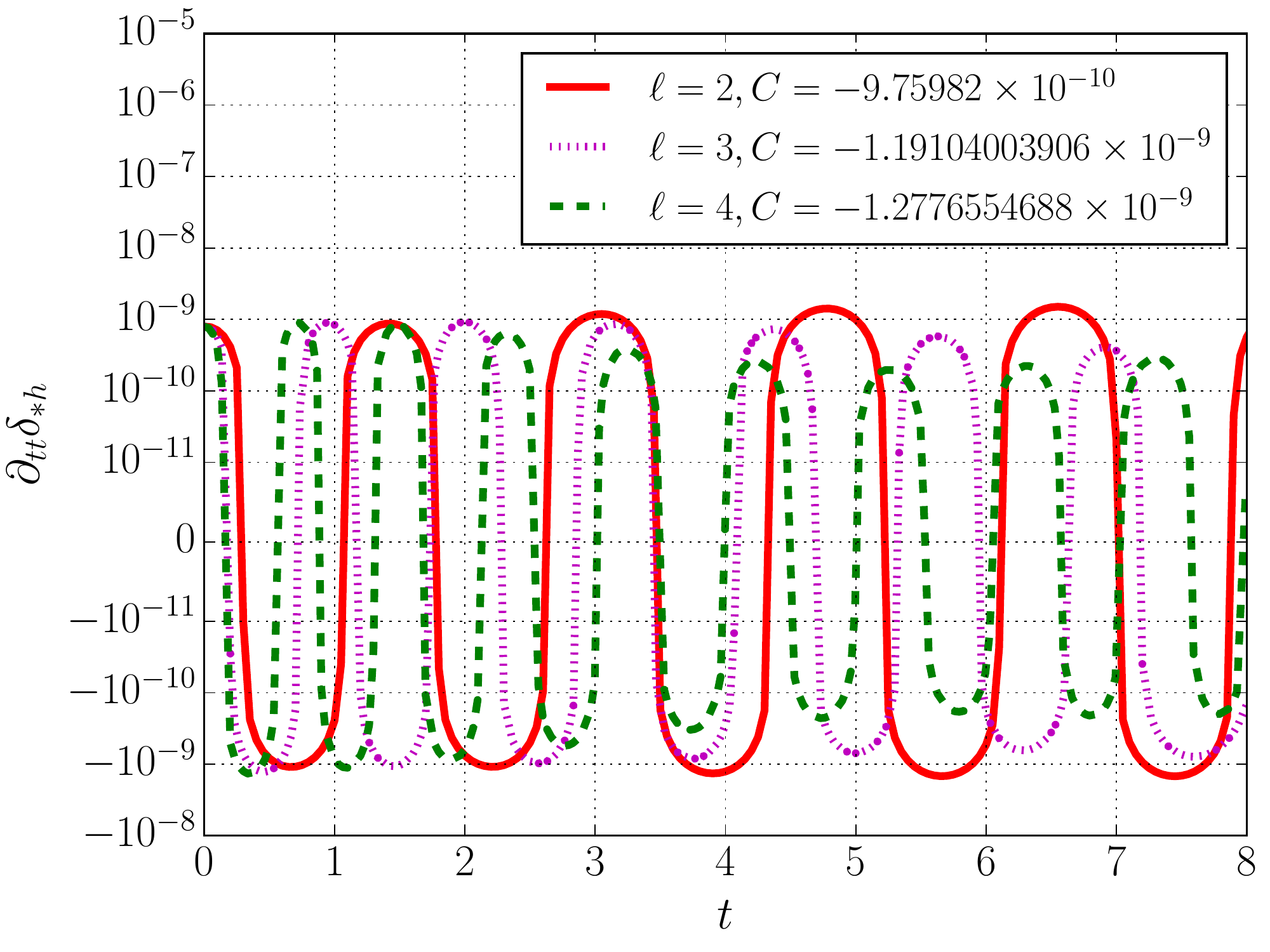}  
     \caption{The fundamental mode of our best numerical approximations of the critical solutions for various values of $\ell$ and fixed $\epsilon=-10^{-4}$.} \label{fig:Different_ls}    
  \end{minipage}  
\end{figure}

Based on the heuristic analysis in \Sectionref{sec:heuristicmode} we now present our numerical findings which support Result~\ref{conj1} in \Sectionref{sec:intro}. Recall from our previous discussion that  the critical behavior is mainly governed by the $l=0$-mode of $\delta_*$. The dynamics of this mode is approximated by \Eqref{eq:linsol31N} which suggests that, for initial data given by any fixed values $\epsilon$ and $\ell\ge 2$, the corresponding solution of the (fully nonlinear) evolution equation should be eventually expanding globally in space if $C<C_{crit}$, where $C_{crit}$ is given by \Eqref{eq:lincritC}, and be eventually collapsing globally in space if $C>C_{crit}$. The critical case should therefore be $C=C_{crit}$. Our numerical results now indeed confirm this, but with a slightly different value $C_{crit}$ than the value in \Eqref{eq:lincritC}. This suggests that there are nonlinear effects in the evolution equations, in particular effects of order $\epsilon^2$, which are not taken into account by our mode analysis (which was based on setting $F_l$ in \Eqref{eq:formallin} to zero). In any case, we find that the actual value is proportional to $\epsilon^2$ in leading order in consistency with \Eqref{eq:lincritC}; cf.\ \Figref{fig:epsilons_and_Cs}.
In practice, we use the following algorithm to determine the actual value of $C_{crit}$ for any choice $\epsilon\in\R$ and $\ell\ge 2$ which is suggested by \Eqsref{eq:linsol31N} -- \eqref{eq:linsol32N}:
\begin{enumerate}
\item Construct the full set of initial data as outlined in \Sectionref{sec:familyIDN} for the given values of $\epsilon$ and $\ell$, and for the value $C$ given by \Eqref{eq:lincritC}.  
\item Evolve the initial data to the future using the fully nonlinear evolution equations and gauges in \Sectionref{sec:evolutionN}. Determine whether the solution collapses (i) or expands (ii) globally in space.
\item Construct new ID in the same way as before for the same value of $\epsilon$ and $\ell$, but with some slightly decreased value of $C$ if (i) in Step 2, or, with some slightly increased value of $C$ if (ii); cf.\ \Eqref{eq:linsol31N}.
\item Go back to Step 2 and repeat this process until a sufficiently good approximation of the critical solution has been obtained.
\end{enumerate}
This algorithm is now used in \Figref{fig:CriticalBehaviour} to approximate the actual fully nonlinear critical solution for $\epsilon=-10^{-4}$ and $\ell=2$. It demonstrates that the time period for which $\delta_{*h}$ is bounded (and oscillatory, see below) is longer the closer $C$ is to the critical value.
In \Figsref{fig:Different_epsilons} and \ref{fig:Different_ls}  we apply the algorithm now to various values of $\epsilon$ and fixed $\ell$, only plotting our best numerical approximation of the critical solution obtained by our algorithm.

\begin{figure}[t]
  \centering
      \includegraphics[width=0.49\textwidth]{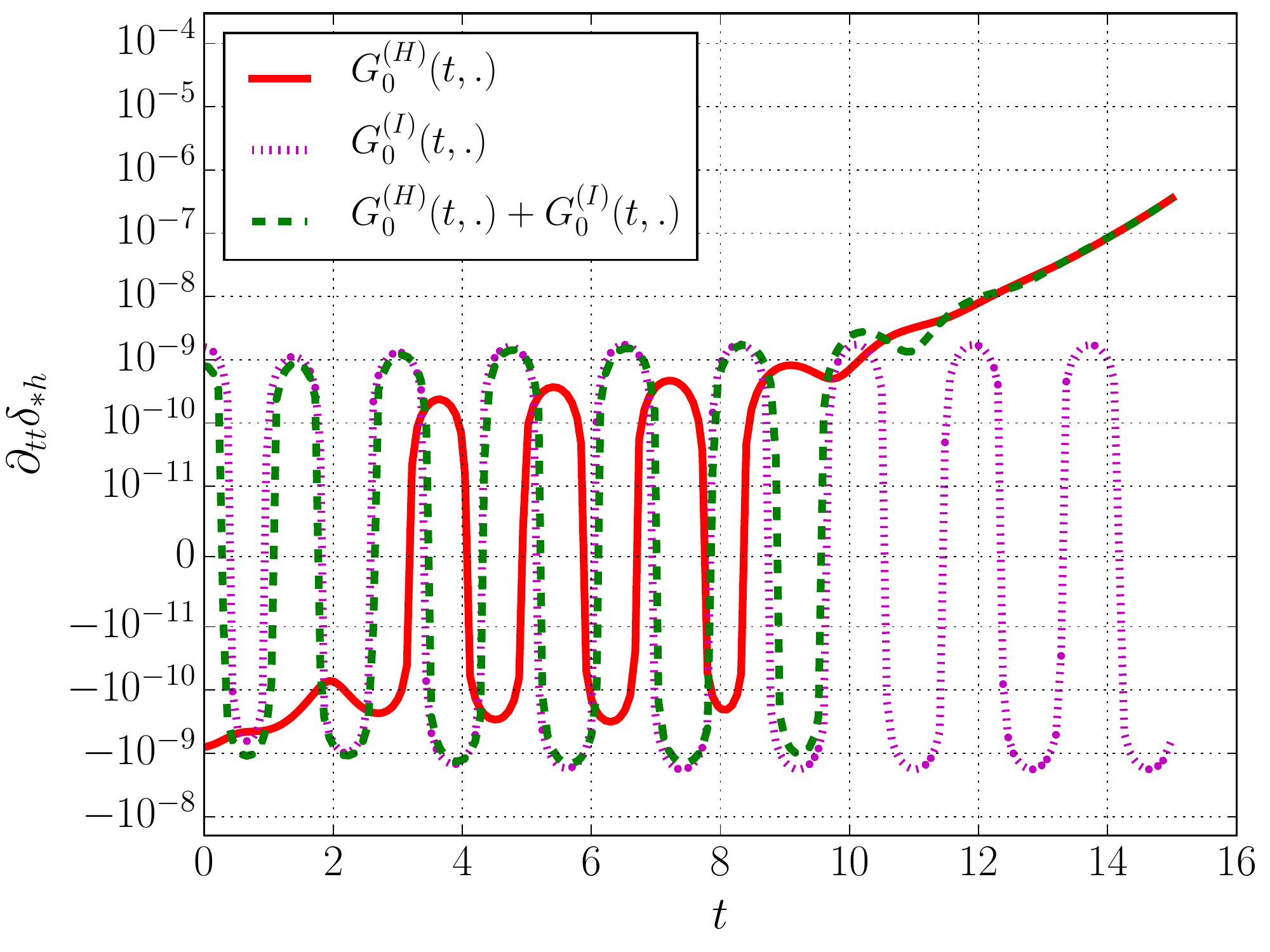}   
     \caption{Homogeneous and inhomogeneous contributions to the evolution of the homogeneous mode as explained in the text for $C= -0.975982 \times 10^{-9}$, $\epsilon= - 10^{-4}$, $\ell= 2$.} \label{fig:Splitting_contributions}
\end{figure}

All these plots \Figsref{fig:CriticalBehaviour}, \ref{fig:Different_epsilons} and \ref{fig:Different_ls} therefore confirm  Result~\ref{conj1} in \Sectionref{sec:intro}. In the next subsection, we shall study the oscillations.
Before we get to this, let us emphasize that none of our numerical solutions is (as a matter of principle) \emph{exactly} critical. Eventually during the evolution, the solutions ``all make a decision whether to expand or to collapse''. 
Let us discuss how this happens in terms of the following alternative decomposition of the evolution equation of the $l=0$-mode
\begin{equation}
  \label{eq:alternativedecomp}
  \ddot\delta_{*,0}(t) = G_0^{(H)}(t, u_{*,0}, \dot u_{*,0} ) + G_0^{(I)}(t, u_{*,1}, u_{*,2},\ldots, \dot u_{*,1}, \dot u_{*,2},\ldots).
\end{equation}
The first term on the right-hand side captures \textit{all} terms (also all nonlinear ones) in the equation which are present in the \textit{spatially homogeneous} case in which we fully understand the criticality of the Nariai solution \cite{Beyer:2009vm}.  The second term can then be considered as ``inhomogeneous corrections'' to the equations close to the homogeneous case. 
Now, \Figref{fig:Splitting_contributions} suggests that the expected unstable behavior is triggered once the homogeneous term $G_0^{(H)}(t, u_{*,0}, \dot u_{*,0} )$ dominates the right-hand side of \Eqref{eq:alternativedecomp} and the evolution therefore displays the well-known Nariai-like instability. The solution plotted there is again our best numerical approximation of the critical solution in \Figref{fig:CriticalBehaviour}. We find that whatever the value of $\delta_{*,0}$ is at the time when the homogeneous term takes over determines whether the solution eventually expands or collapses globally in space.

\subsection{Numerical evidence for Result~2: Nonlinear oscillatory dynamics}
As discussed before, if the solution is critical (or almost critical) and hence $\delta_{*}$ is bounded for an extended period of time, the oscillatory nature of the modes $\delta_{*,l}$ for $l\ge 2$ suggested by \Eqref{eq:linsol32N} should dominate the dynamics. We shall discuss the dynamics of the modes $l\ge 2$ first now, before we explain the oscillatory behavior of the $l=0$-mode in \Figsref{fig:CriticalBehaviour}, \ref{fig:Different_epsilons} and \ref{fig:Different_ls} (which is clearly not explained by \Eqref{eq:linsol31N}).

As before we assume that $\ell\ge 2$.
Recall that the oscillation period of the modes $\delta_{*,l}$ for $l\ge 2$ is expected to be $T_L$ given by \Eqref{eq:linoscillationperiodN}.  In Table~\ref{tab:elloscillations} we confirm the validity and accuracy of this heuristic prediction.
\begin{table}[t]
\centering  
\begin{tabular}{ | c || l | l || l | l |l|l|}
    \hline
    Data       & $t_{L}$      & $t_{N}$    & $T_{L}$   & $T_{N_1}$ & $T_{N_2}$    & $T_{N_3}$  \\  \hline\hline
    $\ell = 2$ & $0.723221$   &  $0.69999$ & $3.14159$ & $3.19999$& $3.5499997$   &  ...       \\  \hline
    $\ell = 3$ & $0.47418 $   &  $0.46662$ & $1.98692$ & $2.09979$& $2.3997598$& $2.3997598$ \\  \hline
    $\ell = 4$ & $0.359534$   &  $0.333$   & $1.48096$ & $1.5651$ & $1.7649001$& $2.1644998$ \\  \hline 
\end{tabular}
\caption{Critical solutions for $\epsilon=-10^{-4}$ and various $\ell$: Oscillatory behavior of $\delta_{*,\ell}$. Here, $T_L$ is the prediction from \Eqref{eq:linoscillationperiodN}. $T_{N_1}$ is the actual period length of the first full oscillation after $t=0$, and $T_{N_2}$ and $T_{N_3}$ of the second and third one. Moreover, $t_L$ is the predicted time of the first oscillation maximum according to \Eqref{eq:linsol32N} and $t_N$ the actual numerical value.}
\label{tab:elloscillations}
\end{table}
As expected the agreement with \Eqref{eq:linsol32N} is better for smaller $t$ and indeed gets worse for the second and third oscillation when non-linear effects clearly become significant.

\begin{figure}[t]
  \begin{minipage}{0.49\linewidth}
    \centering
     \includegraphics[width=\textwidth]{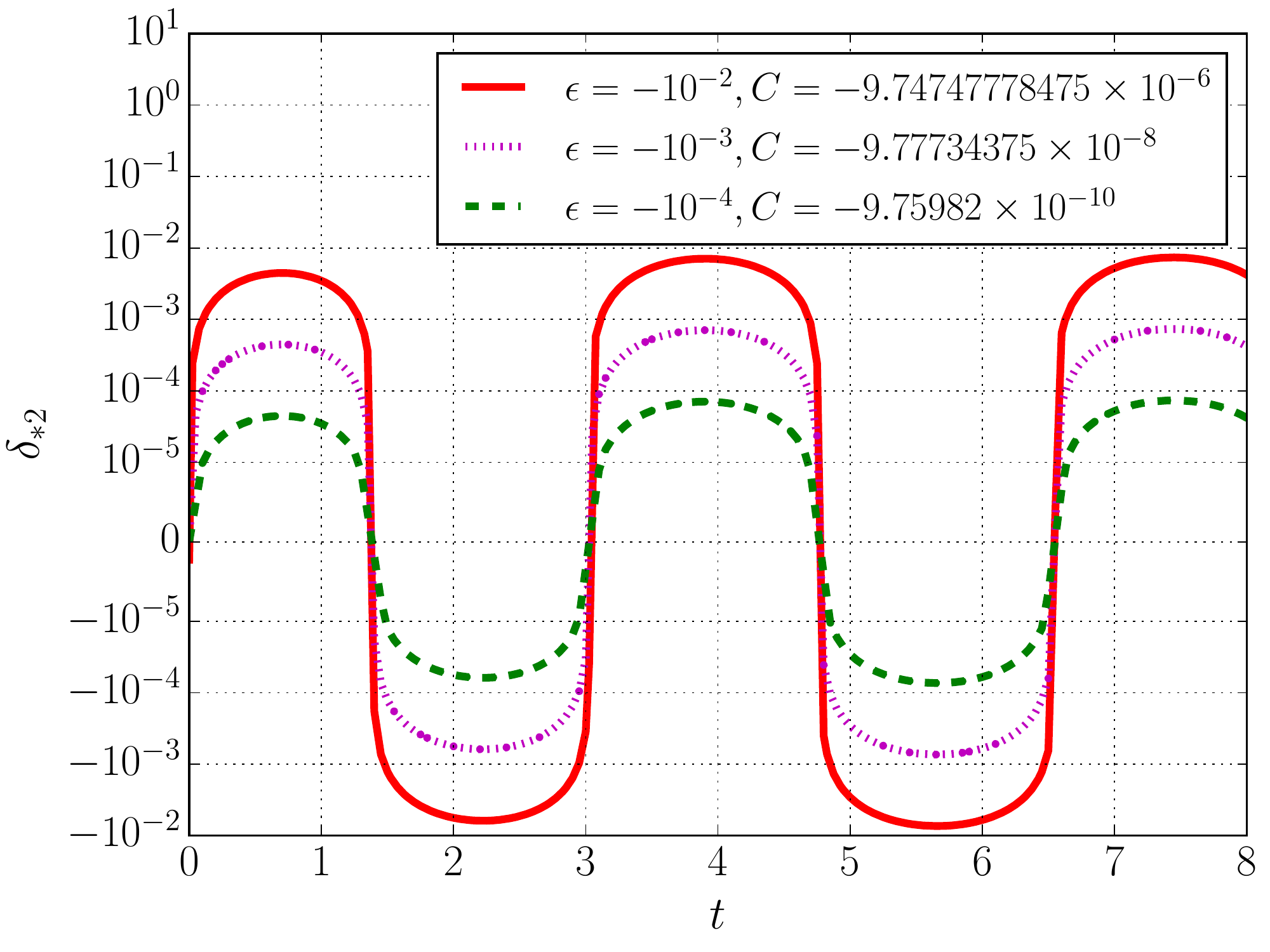}  
     \caption{The $l=\ell$-mode of our best numerical approximations of the critical solutions for various values of $\epsilon$ and fixed $\ell=2$.}
     \label{fig:ell2l2}    
  \end{minipage}  
  \hfill
  \begin{minipage}{0.49\linewidth}
    \centering
     \includegraphics[width=\textwidth]{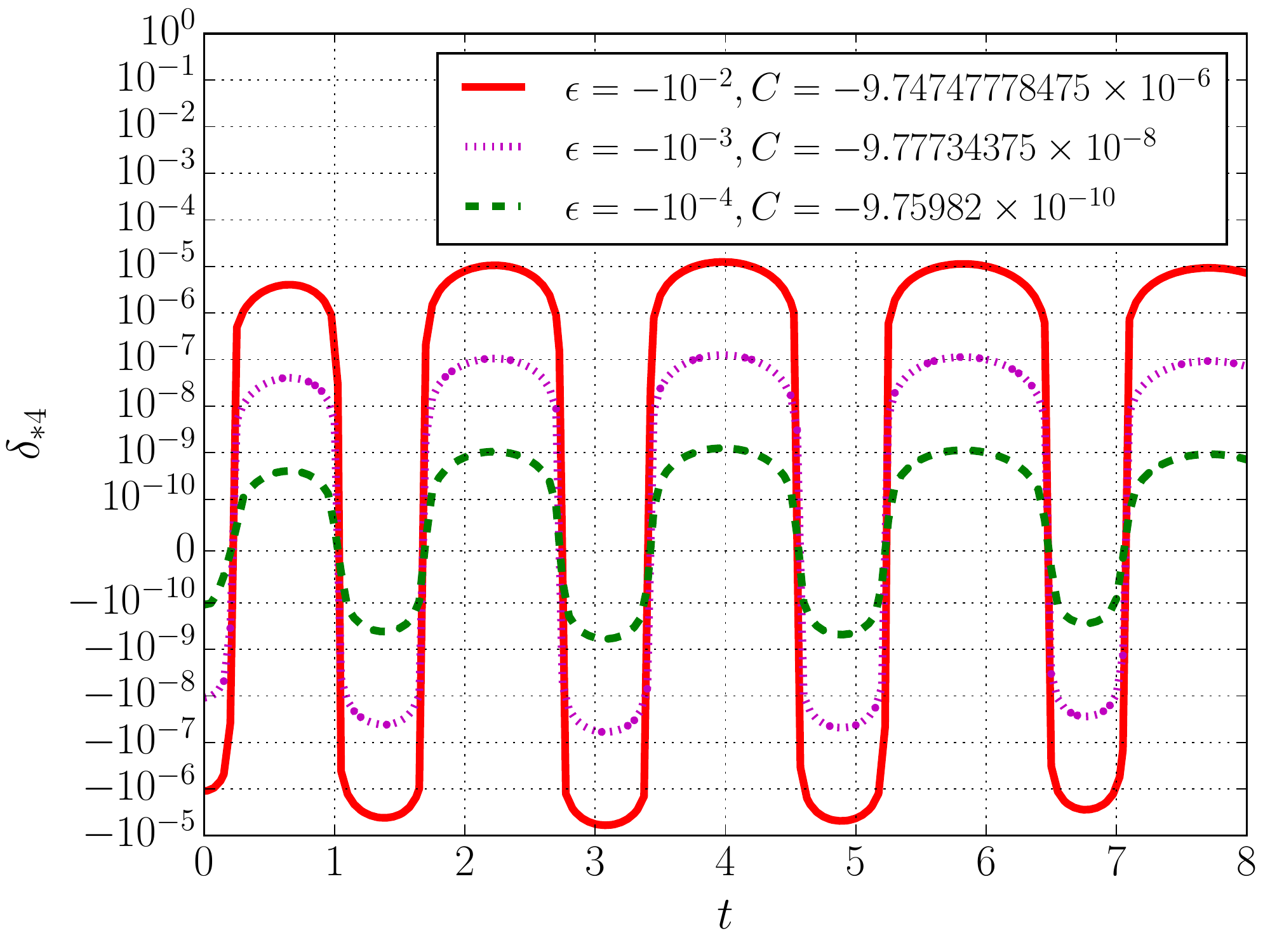}  
     \caption{The $l=4$-mode (i.e., $l\not=\ell$) of our best numerical approximations of the critical solutions for various values of $\epsilon$ and fixed $\ell=2$.}
     \label{fig:ell2l4}  
  \end{minipage}  
\end{figure}

Recall from \Eqref{eq:linsol32N} that there should be a phase difference of $\pi/2$ between the oscillations of the modes with $l\not=\ell$ ($l=2,\ldots,2\ell$), and the mode $l=\ell$. Moreover, the amplitudes of all the former modes should be proportional to $\epsilon^2$ while the amplitude of the latter is proportional to $\epsilon$. All this is confirmed in \Figsref{fig:ell2l2}  and \ref{fig:ell2l4}.  
We also remark here that \Figref{fig:Different_epsilons} confirms, in consistency with the heuristic predictions, that the oscillation period only depends on $\ell$ but not on $\epsilon$ or $C$.

Now, while our heuristic analysis explains that the fundamental mode is bounded for the (almost) critical solutions and all modes $l\ge 2$ are oscillatory, it misses the oscillatory behavior of the fundamental mode which is obvious in \Figsref{fig:CriticalBehaviour}, \ref{fig:Different_epsilons} and \ref{fig:Different_ls}.
The basic assumption for the results in \Sectionref{sec:heuristicmode} was that the term $F_l$ in \Eqref{eq:formallin} is negligible. For the fundamental mode, this is clearly not the case and terms which are $O((\delta_{*,\ell})^2+(\partial_t\delta_{*,\ell})\delta_{*,\ell}+(\partial_t\delta_{*,\ell})^2)$ are expected to change the dynamics significantly; recall that the amplitudes of all modes $\delta_{*,l}$ with $l\not=\ell$ are of order $\epsilon^2$ and hence of higher order than $\delta_{*,\ell}$. Because of the quadratic coupling of the fundamental mode and the  $l=\ell$-mode, whose amplitude is proportional to $\epsilon$ and whose oscillation period length is approximately given by \Eqref{eq:linoscillationperiodN},  we expect that the amplitude of the oscillation of the fundamental mode is proportional to $\epsilon^2$ and the oscillation period length is half of \Eqref{eq:linoscillationperiodN}, i.e.,
\begin{equation}
  \label{eq:linoscillationperiodnonlin}
  \frac{\pi}{\sqrt{\ell(\ell+1)-2}}.
\end{equation}
The statement about the amplitude is indeed confirmed by \Figref{fig:Different_epsilons}. The accuracy of the prediction about the oscillation period length is studied in Table~\ref{tab:elloscillations2}.
\begin{table}[t]
\centering  
\begin{tabular}{ | l || l | l | l | l |}
    \hline
    Data       &  $T_{L}$      & $T_{N_1}$  & $T_{N_2}$ & $T_{N_3}$\\    \hline\hline
    $\ell = 2$ & $1.5708$  & $1.65$&   $1.7500002$ & $1.75$ \\   \hline
    $\ell = 3$ & $0.993459$& $1.1665499$& $1.1105$ & $0.9665699$ \\   \hline
    $\ell = 4$ & $0.74048$ & $0.7992$   & $0.86580002$ & $0.73259997$ \\   \hline 
\end{tabular}
\caption{Critical solutions for $\epsilon=-10^{-4}$ and various $\ell$: Oscillatory behavior of $\delta_{*,0}$. Here, $T_L$ is the prediction from \Eqref{eq:linoscillationperiodnonlin}. $T_{N_1}$ is the actual period length of the first full oscillation after $t=0$, and $T_{N_2}$ and $T_{N_3}$ of the second and third one.}
\label{tab:elloscillations2}
\end{table}

\begin{figure}[t]
  \centering
  \begin{minipage}{0.49\linewidth}
    \centering
    \includegraphics[width=\textwidth]{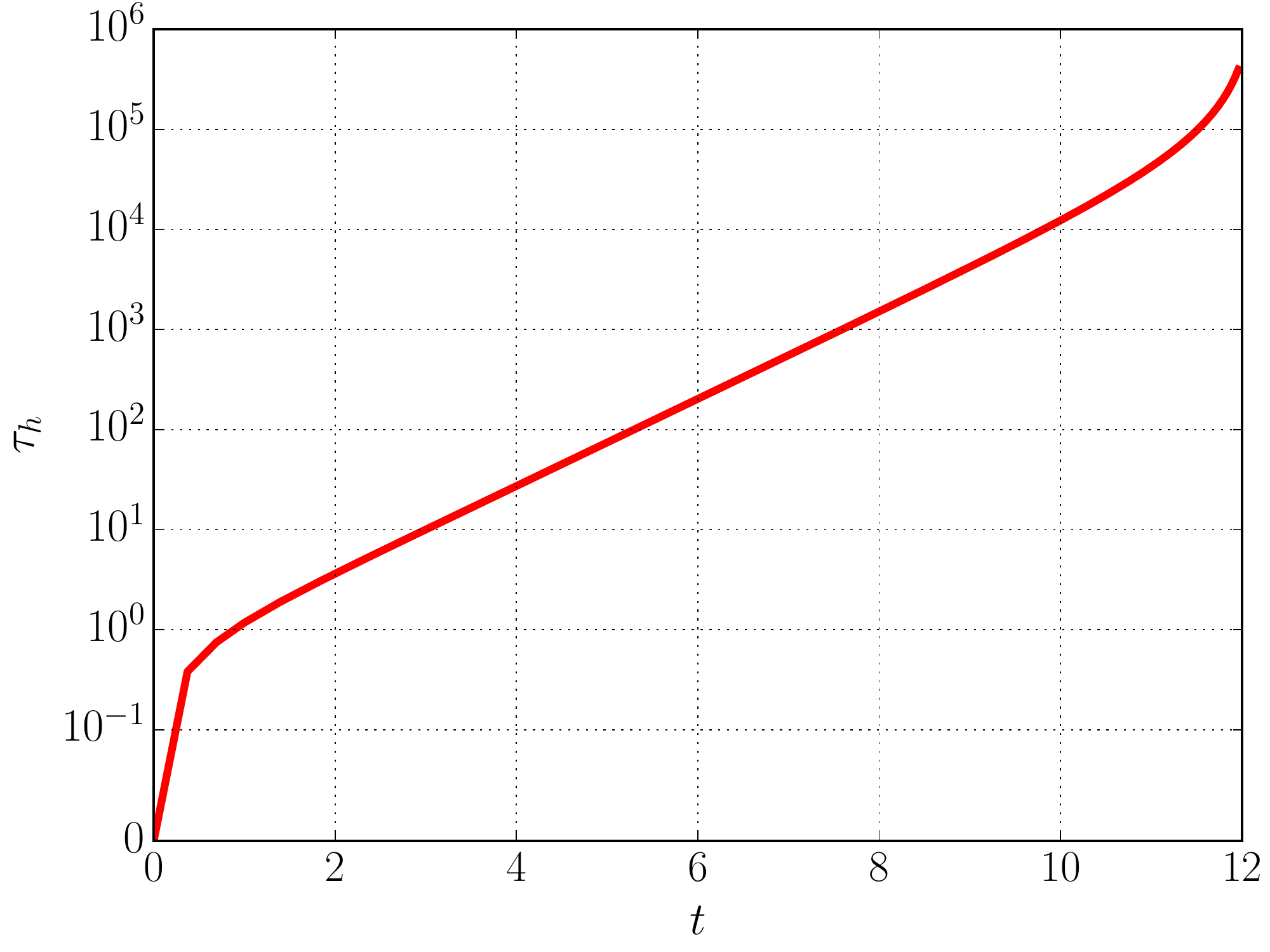}
    \caption{Fundamental mode of the solution of the Eikonal equation discussed in the text for the solution given by $C= 10^{-4}$, $\epsilon= - 10^{-4}$,
      $\ell= 2$.}
    \label{fig:ProperTimeN}
  \end{minipage}   
  \hfill
  \begin{minipage}{0.49\linewidth}
    \centering
    \includegraphics[width=\textwidth]{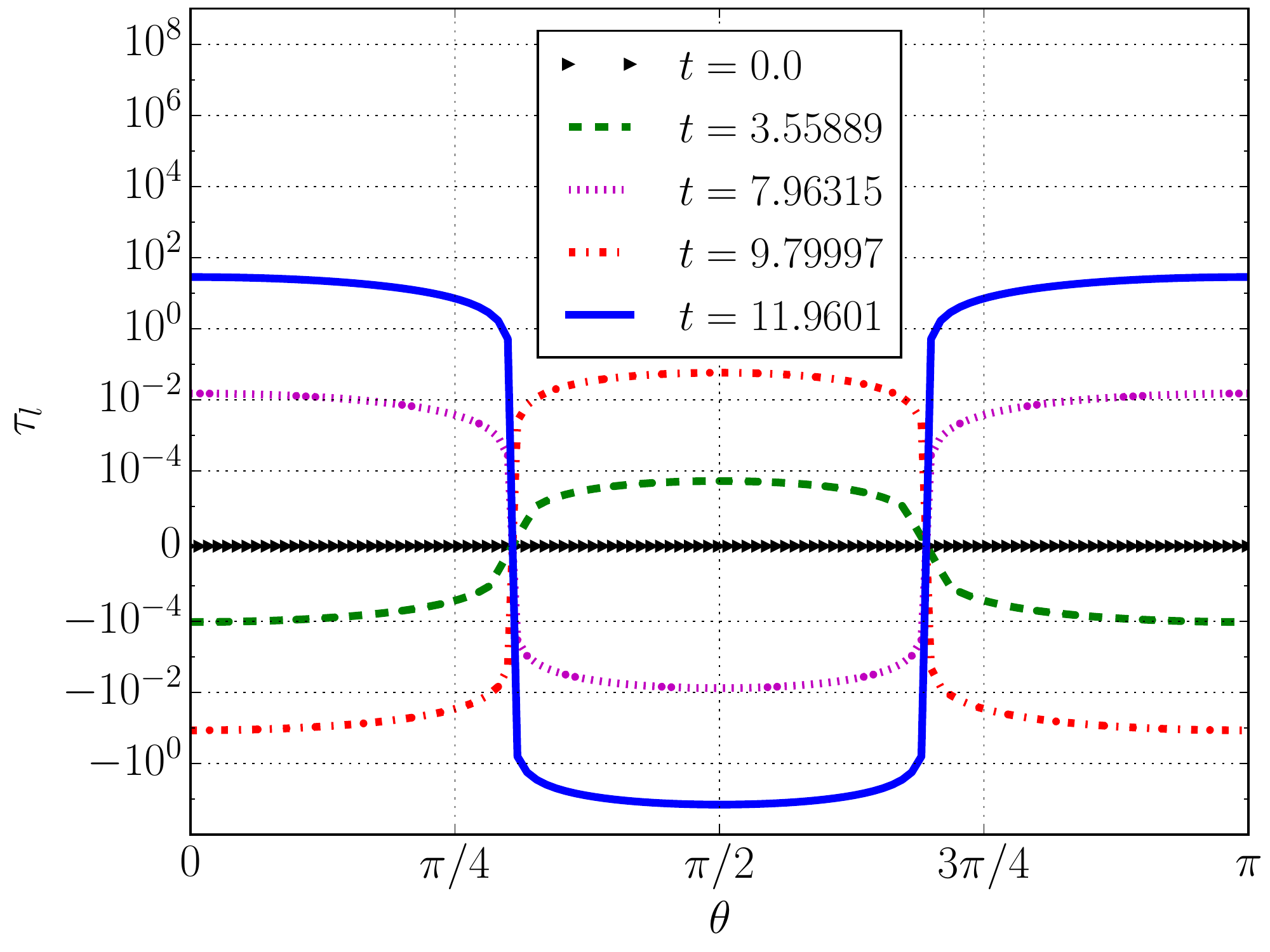}
    \caption{Inhomogeneous part of the solution of the Eikonal equation discussed in the text for the solution given by $C= 10^{-4}$, $\epsilon= - 10^{-4}$,
      $\ell= 2$, i.e., $\tau_l=\tau-\tau_h$.}
    \label{fig:ProperTime2}
  \end{minipage}   
\end{figure}

\begin{figure}[t]
     \centering
     \includegraphics[width=0.49\textwidth]{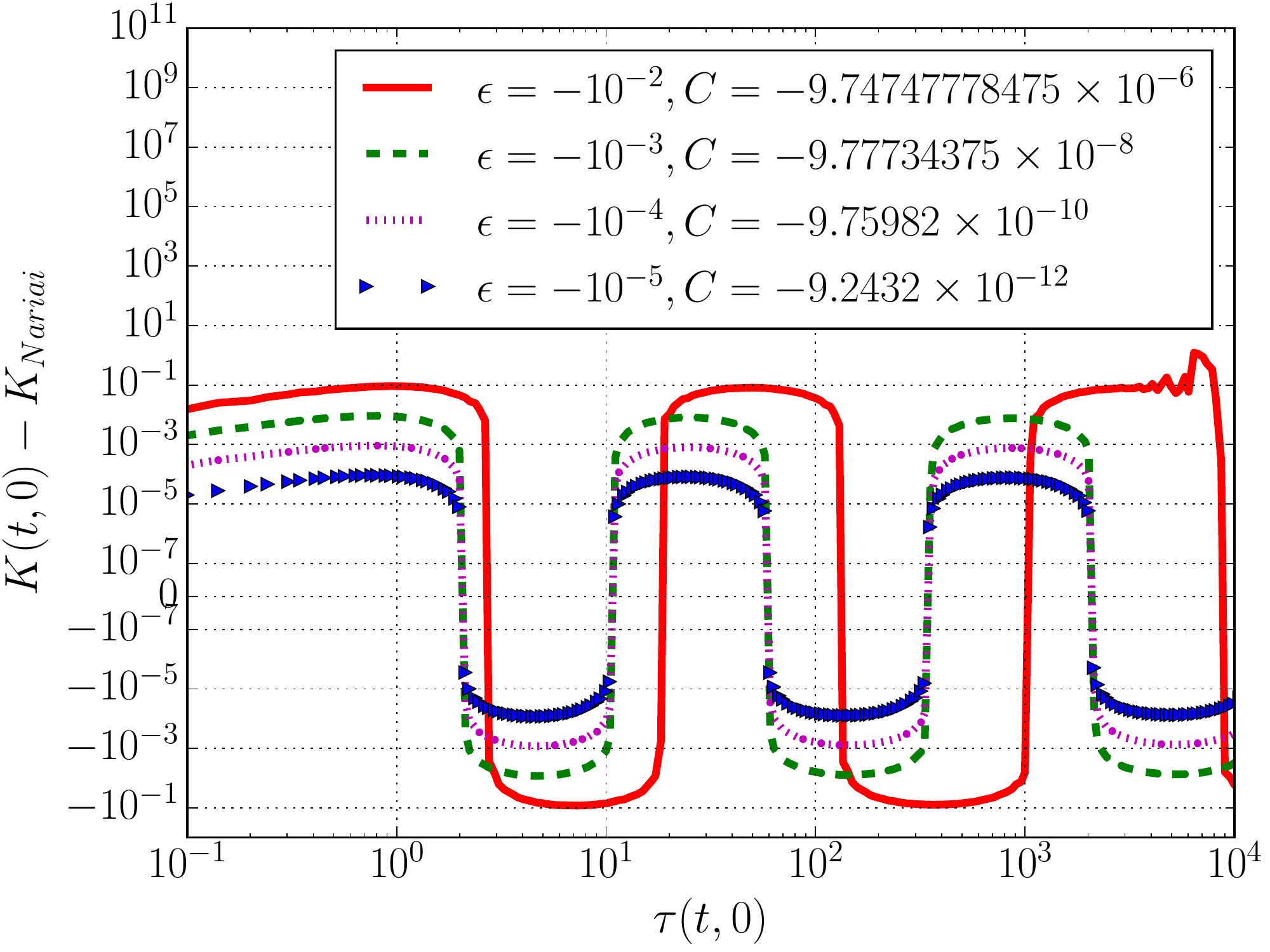}
     \caption{Kretschmann scalar $K$ of the $3+1$-metric vs.\ the physical time $\tau$ at the pole $\theta=0$ for various best approximations of critical solutions for $\ell=2$.}
     \label{fig:KretschmannOscillations}
 \end{figure}

In the rest of this subsection we consider the question whether these oscillations are a real physical effect of our models rather than just a gauge effect. To this end, we define a physical time function as the solution $\tau$ of the Eikonal equation
\begin{equation}\label{eq:eiconal}
  \nabla_a \tau \nabla^a \tau = -1
\end{equation}
with zero initial data on any of our models. As explained in \cite{Beyer:2016fc}, the value of $\tau$ represents the proper time along the congruence of unit timelike geodesics which start perpendicularly to the initial hypersurface. 
In \Figsref{fig:ProperTimeN} and \ref{fig:ProperTime2}, we plot this function $\tau$ in one of our cosmological models, mainly to demonstrate that our numerical solutions cover a significant part of \textit{physical time}.
Let us now describe our oscillations in terms of the physical time before. In order to simplify the discussion a little, we exploit the fact that for $3+1$-Gowdy symmetric spacetimes and hence for axi-symmetric $2+1$-spacetimes, the poles of the spatial two-sphere is a geometrically distinguished point at all times. It is therefore geometrically (and physically) meaningful to look at the  Kretschmann scalar of the $3+1$-metric as a function of $\tau$ at the pole $\theta=0$ of the spatial $2$-spheres only. This is done in \Figref{fig:KretschmannOscillations} for various critical solutions. The oscillations are evident in this representation and hence are a real physical phenomenon

All this confirms Result~\ref{conj2} in \Sectionref{sec:intro}.

\subsection{Numerical evidence for Result 3: Late time behavior}
\begin{figure}[t]
  \centering
     \includegraphics[width=0.49\textwidth]{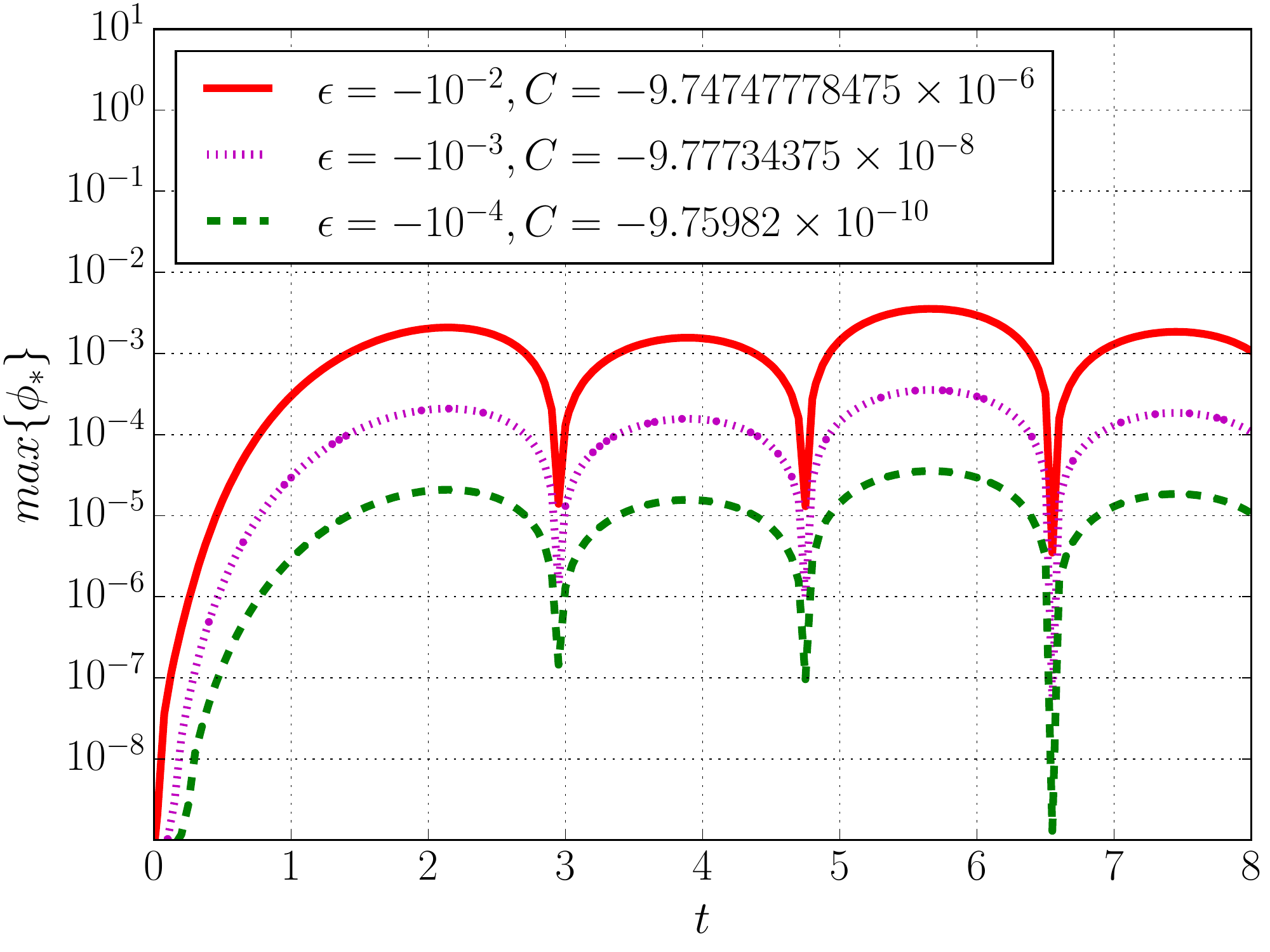}
     \caption{The quantity $\phi_*$ for our best numerical approximations of the critical solutions for various values of $\epsilon$ and fixed $\ell=2$.}     
     \label{fig:phiOscillations}
 \end{figure}

\begin{figure}[t]
  \begin{minipage}{0.49\linewidth}
    \centering     
      \includegraphics[width=\textwidth]{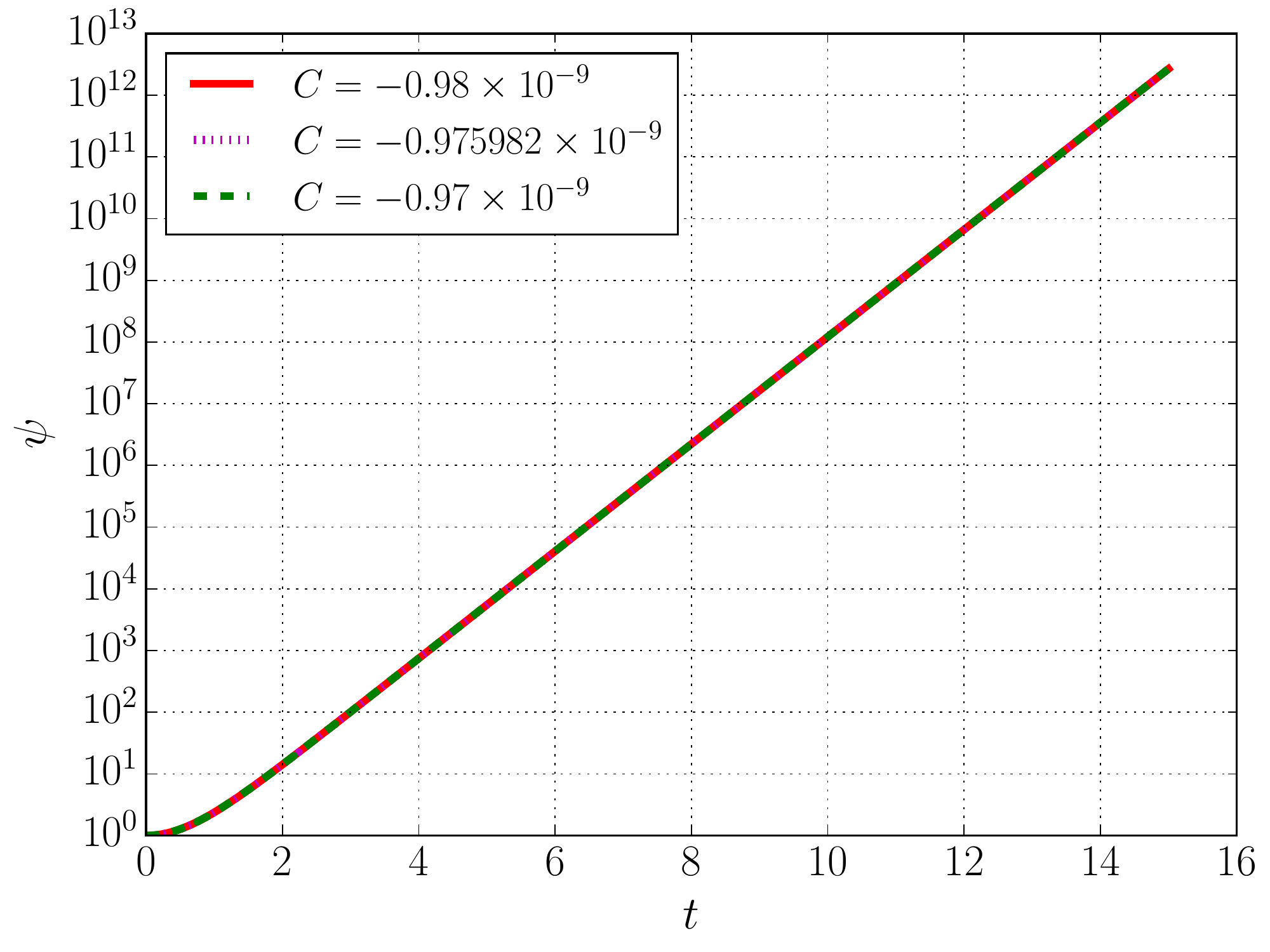}  
     \caption{Exponential growth of $\psi$ for models given by $\epsilon=10^{-4}$, $\ell=2$ and various values of $C$ (close to the critical value).} \label{fig:psi1}
  \end{minipage}  
   \hfill    
     \begin{minipage}{0.49\linewidth}
    \centering
     \includegraphics[width=\textwidth]{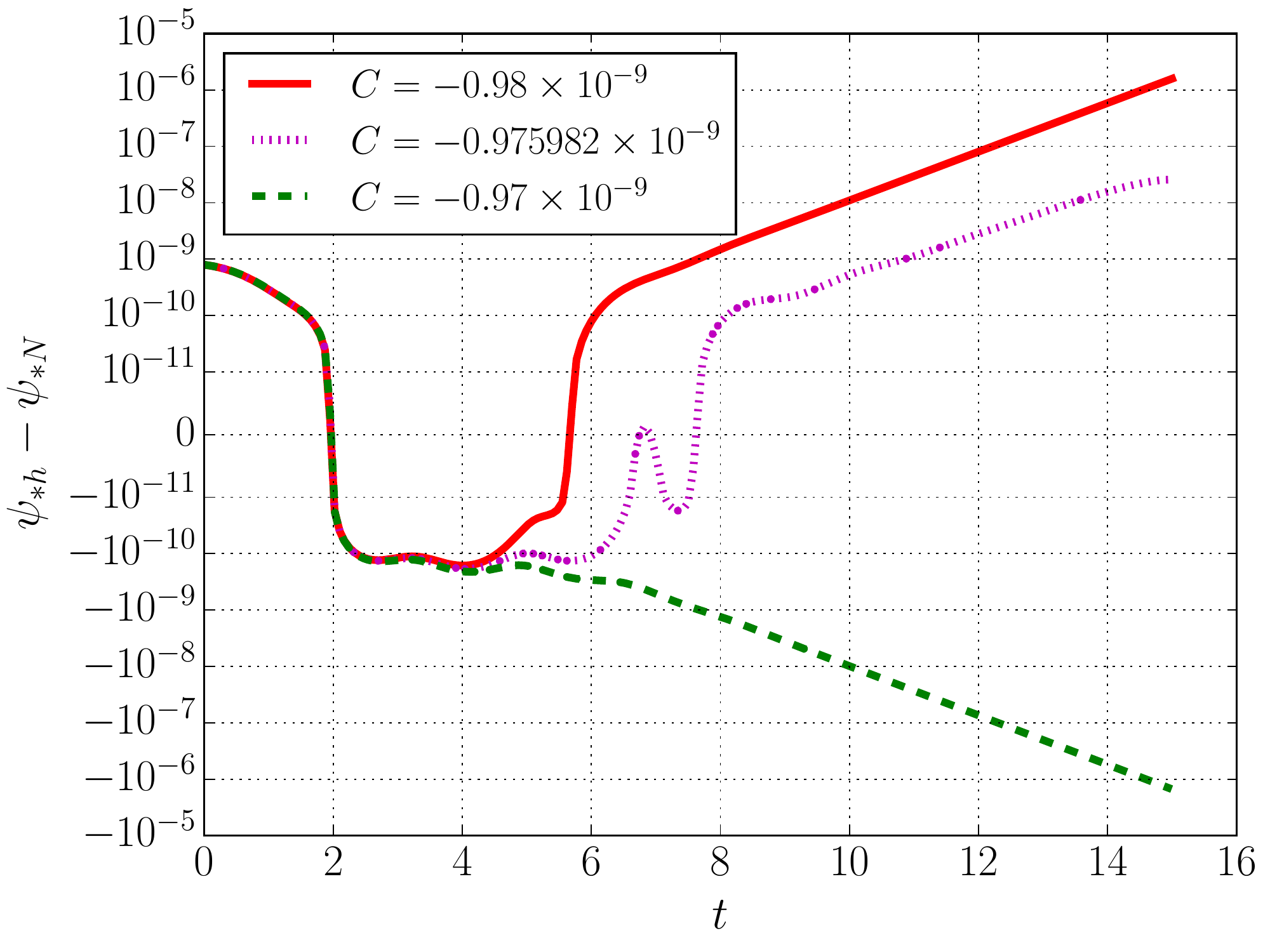}  
     \caption{The difference of $\psi$ and the corresponding quantity of the Nariai solution is small for $\epsilon=10^{-4}$, $\ell=2$ and various values of $C$ (close to the critical value). } \label{fig:psi2}    
  \end{minipage}  
\end{figure}

We have now used numerical evidence to support our claim that for any $\epsilon\in\R$ and $\ell\ge 2$  it is possible to keep the quantity $\delta_*$ bounded for as long as we like by picking $C$ sufficiently close to some critical value.
\Figref{fig:phiOscillations} now shows that also the quantity $\phi_*$ (see the definition in \Eqref{eq:defdeltaS}) is bounded and, in fact, oscillatory. Since $\delta_*$ and $\phi_*$ determine the physical geometry of the spatial two-spheres (recall \Eqref{eq:physmetric}), it follows that the spatial $\St$-factor of critical $3+1$-models do not deviate much from the geometry of the standard round unit $2$-sphere  for an arbitrary long time, on the one hand.
On the other hand, the geometry of the spatial $\So$-factor, which is described by the quantity $\psi$,  changes exponentially, as suggested by \Figref{fig:psi1}, and its growth is almost unaffected by whether $C$ is larger or smaller than the critical value.  See also \Figref{fig:psi2} which shows the deviation of this quantity from the corresponding Nariai values. Recall that $\psi_*$ is defined in \Eqref{eq:defpsiS}. 

\begin{figure}[t]
  \begin{minipage}{0.49\linewidth}
    \centering     \includegraphics[width=\textwidth]{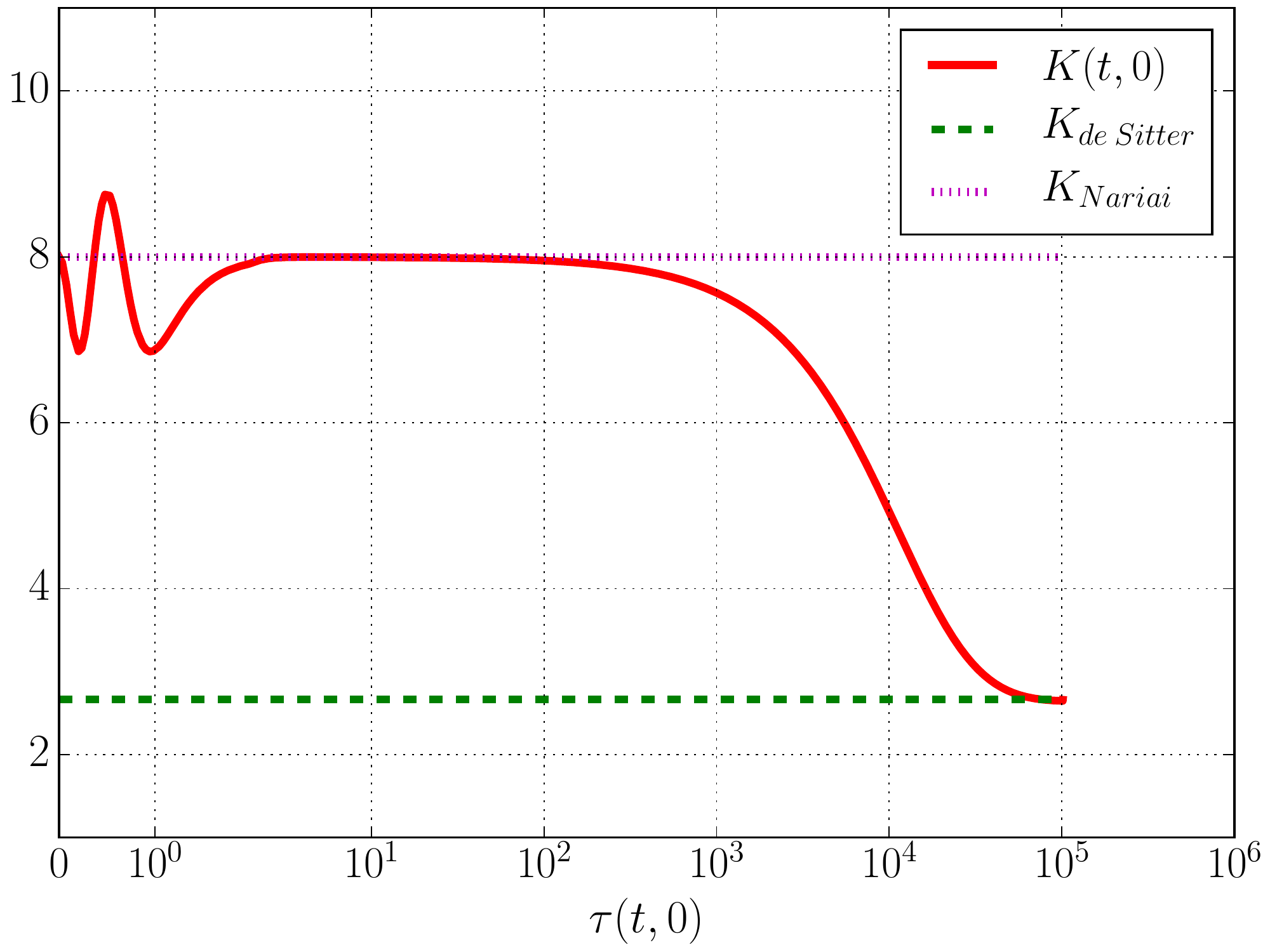}               
    \caption{The value of the Kretschmann scalar at the pole vs.\ the
      value of $\tau$ at the pole. Late time behavior for the solution given by $C= -10^{-4}$, $\epsilon= - 10^{-4}$,   $\ell= 2$.}\label{fig:Kretschmann_Homo_part}  
  \end{minipage}  
   \hfill    
     \begin{minipage}{0.49\linewidth}
    \centering
     \includegraphics[width=\textwidth]{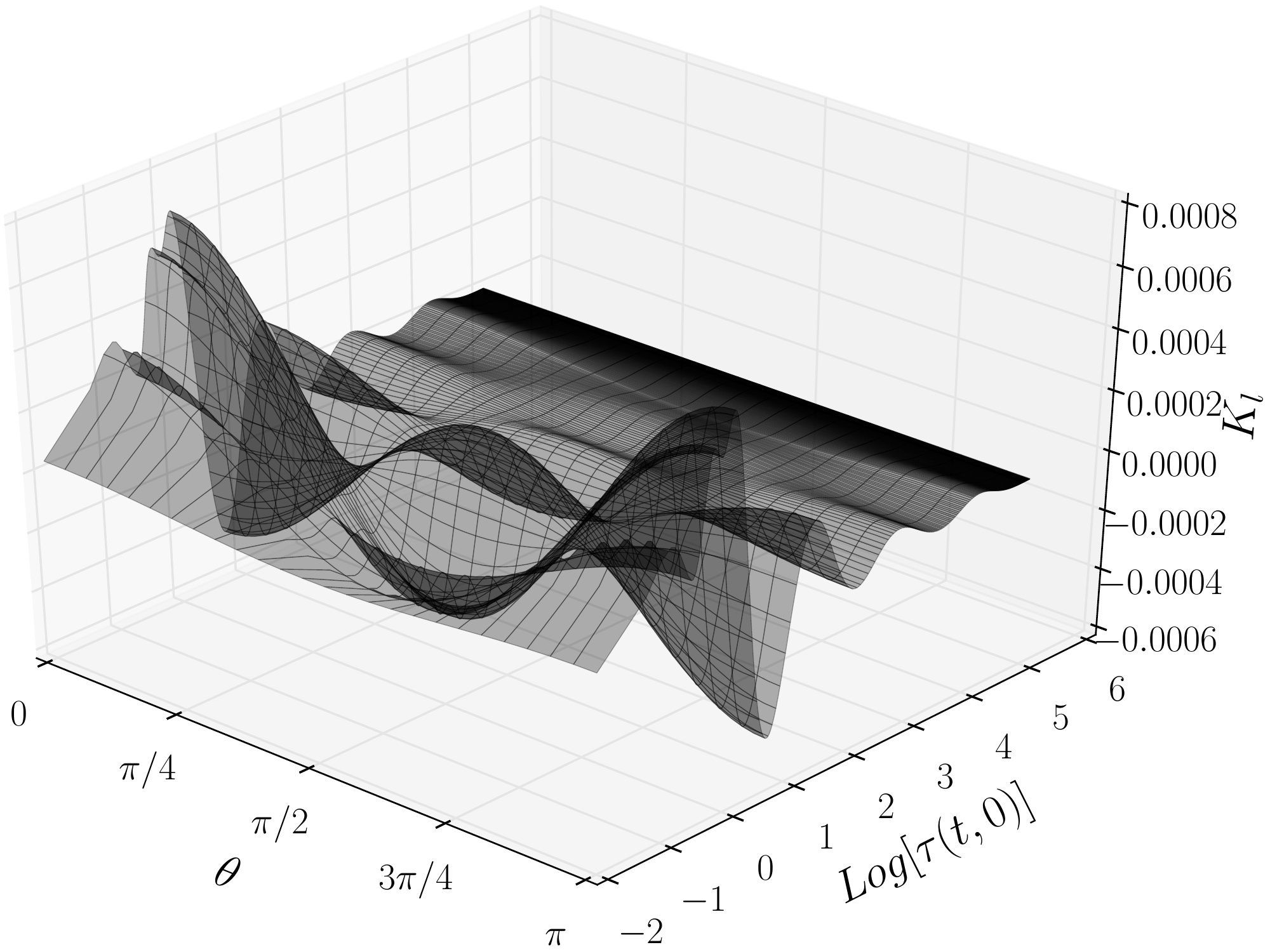}  
     \caption{Evolution of $K_l = K-K_h$ (where $K$ is the Kretschmann scalar) vs.\ the
      value of $\tau$ at the pole. Late time behavior for the solution given by $C= -10^{-4}$, $\epsilon= - 10^{-4}$,   $\ell= 2$. \\} \label{fig:Kretschmann_Harmonic_part}    
  \end{minipage}  
\end{figure}

This supports the claim that the long-term behavior of our inhomogeneous critical solutions is very similar to that of the exact Nariai solution. In particular, as for the Nariai solution, the highly anisotropic timelike future is expected to be inconsistent with the cosmic no hair picture. This supports Result~\ref{conj3} in \Sectionref{sec:intro}. As explained earlier, we have convincing evidence now that whenever the solutions are non-critical, they either expand or collapse to the future eventually. In the expanding case, we expect the solutions to behave {in accordance} with the cosmic no-hair conjecture. This is indeed confirmed by our numerical results. For instance, \Figsref{fig:Kretschmann_Homo_part} and \ref{fig:Kretschmann_Harmonic_part}  show (for a clearly non-critical solution; the value $C=-10^{-4}$ is far in the expanding regime) that, while  the $3+1$-Kretschmann scalar starts off close to the Nariai value, it eventually approaches the expected de-Sitter value after all the oscillations have died out.

%%%%%%%%%%%%%%%%%%%%%%%%%%%%%%%%%%%%%%%%%%%%%%%%%%%%%%%%%%%%%%%%%%%%%%%%%%%%%%%%%%%%%%%%%%%%%%%%%%

\section*{Acknowledgments}
J.~F.\ would like to thank the Department of Mathematics at the University of Oslo for hospitality. Part of this research was supported by the European Research Council through the FP7-IDEAS-ERC Starting Grant scheme, Project No. 278011 STUCCOFIELDS.
The author L.\ E.\ was partly funded by the University of Otago Research Grant ``Dynamical dark energy in the young universe and its consequences for the present and future history'' in 2016. We would like to thank Dr Chris Stevens for helpful discussions. 

\appendix

\section{Spin-weighted spherical harmonics}\label{Sec:swsh}

Let $(\theta,\varphi)$ be the standard polar coordinates in $\St$. A function $f$ on $\mathbb{S}^2$ has spin-weight $s$ if it transforms under a local rotation by an angle $\tau$ in the tangent plane at every point $(\theta,\varphi) \in \St$  as $f \to e^{i s \tau} f$. In this case, $f$ can be written as
\begin{equation}\label{eq:functionS2}
f(\theta,\varphi) =  \sum\limits_{l=|s|}^{\infty}  \sum\limits_{m=-l}^{l} a_{lm}\, {}_{s}Y_{lm} (\theta,\varphi) ,
\end{equation}
where $_{s}Y_{lm}( \theta , \varphi)$ are the spin-weighted spherical harmonics \cite{Penrose:1984tf} and $a_{lm}$ are complex numbers. These functions are normalized as
\begin{equation}\label{integral_properties_spherical_harmonics}
 \int \limits_{\mathbb{S}^2} \s  {}_{s} Y_{l_1 m_1 }(\theta,\varphi) \: _{s}\overline{Y}_{l_2 m_2}(\theta,\varphi) \s d\Omega = \delta_{l_1 l_2} \delta_{m_1 m_2}.
\end{equation}
For any function $f$ of spin-weight $s$, this identity can be used to calculate the complex coefficients $a_{lm}$ in \Eqref{eq:functionS2}.

The  \textit{eth} operators $\eth$ and $\bar{\eth}$  are  defined by 
\begin{equation}\label{eq:def_eths}
\begin{split}
\eth f       &:= \partial_\theta f - \dfrac{i}{ \sin \theta} \partial_\varphi f- s f \cot \theta=\sqrt{2} m^a\nabla_a f- s f \cot\theta, \\ 
\bar{\eth} f &:= \partial_\theta f + \dfrac{i}{ \sin \theta} \partial_\varphi f + s f \cot \theta=\sqrt{2} \mbar^a\nabla_a f+ s f \cot\theta  ,
\end{split}
\end{equation}
cf.\ \Eqref{eq:referenceframe},
for any function $f$ on $\mathbb{S}^2$ with spin-weight $s$. 
Using  \Eqsref{eq:def_eths}, we can therefore express the  frame vectors $(m^a,\mbar^a)$ in \Eqref{eq:referenceframe} in terms of the eth-operators as
\begin{equation}\label{eq:ethm}
m^a( f ) = \dfrac{1}{\sqrt 2} \left(  \eth f  + f  s \cot\theta  \right), \quad \mbar^a( f ) = \dfrac{1}{\sqrt 2} \left( \bar{\eth} f  -  f s \cot\theta   \right) .
\end{equation}
The  properties of raising and lowering spin are
\begin{eqnarray}\label{eq:eths}
\eth  \hspace{0.1cm}_{s}Y_{lm} (\theta,\varphi)  &=& - \sqrt{ (l-s)(l+s+1) } \hspace{0.1cm}_{s+1}Y_{lm} (\theta,\varphi) , \nonumber \\
\bar{\eth}   \hspace{0.1cm}_{s}Y_{lm} (\theta,\varphi)   &=& \sqrt{ (l+s)(l-s+1) } \hspace{0.1cm}_{s-1}Y_{lm} (\theta,\varphi) , \\
\bar{\eth} \eth  \hspace{0.1cm}_{s}Y_{lm} (\theta,\varphi)   &=& - (l-s)(l+s+1) \hspace{0.1cm}_{s}Y_{lm} (\theta,\varphi) .\nonumber
\end{eqnarray}
In fact, we can use the relations \eqref{eq:eths} to define spin-weighted spherical harmonics $_{s}Y_{lm}$ with any integer spin-weight $s$ from the standard spherical harmonics
\begin{equation}
  \label{eq:zerospin}
  Y_{lm}(\theta,\varphi)=\,_{0}Y_{lm}(\theta,\varphi).
\end{equation}
It is easy to check that from any function $f$ with spin-weight $s$, we can obtain a function with either spin $s+1$ from $\eth(f)$ or spin $s-1$ from $\bar{\eth}(f)$. Thus, they are also known in the literature as the raising and lowering operators \cite{McEwen:2011ib}. We can also check that
\begin{equation}\label{conmutator_eths}
  [  \bar{\eth} , \eth  ] f = 2 s f.
\end{equation}
The Laplace operator (in our sign convention) of the two-sphere can be written in terms of eth-operators as 
\begin{equation}
  \label{eq:DefDelta}
   \Delta_{\St} f = \dfrac{ \left( \eth   \bar{\eth} +  \bar{\eth}  \eth  \right) }{2}  f \;.
\end{equation}
Further, using the commutation relation \Eqref{conmutator_eths} we obtain the  useful expressions 
\begin{eqnarray}
 \Delta_{\St} f = \eth   \bar{\eth}f + s f = \bar{\eth}  \eth f - s f \; \label{eth_for_constraint1} .
\end{eqnarray}

Finally, any function $f$ with spin-weight $s$ can be expanded in terms of \keyword{axi-symmetric} spin-weighted spherical harmonics
\[_{s} Y_l(\theta)=_{s} Y_{l0}(\theta,\varphi)\]
since the latter is independent of $\varphi$, i.e., \Eqref{eq:functionS2} becomes
\begin{equation}
f(\theta) =  \sum\limits_{l=|s|}^{\infty}  a_{l}\, {}_{s}Y_{l} (\theta),
\end{equation}
for complex numbers $a_l$. In analogy to \Eqref{eq:zerospin}, we shall often write 
\begin{equation}
%  \label{eq:zerospin}
  Y_{l}(\theta)=\,_{0}Y_{l}(\theta).
\end{equation}

%%%%%%%%%%%%%%%%%%%%%%%%%%%%%%%%%%%%%%%%%%%%%%%%%%%%%%%%%%%%%%%%%%%%%%%%%%%%%%%%%%%%%%%%%%%%%%%%%%%%%%%%%%%%%%%%%%%%%%%%%%%%%%%%%

\addcontentsline{toc}{section}{References}
%\bibliography{bibliography}

\end{document}